\PassOptionsToPackage{unicode}{hyperref}
\PassOptionsToPackage{hyphens}{url}
\PassOptionsToPackage{dvipsnames,svgnames*,x11names*}{xcolor}
\documentclass[
  astrosymb, twocolumn, tighten]{aastex63}
\usepackage{lmodern}
\usepackage{amssymb,amsmath}
\usepackage{ifxetex,ifluatex}
\ifnum 0\ifxetex 1\fi\ifluatex 1\fi=0 
  \usepackage[T1]{fontenc}
  \usepackage[utf8]{inputenc}
  \usepackage{textcomp} 
\else 
  \usepackage{unicode-math}
  \defaultfontfeatures{Scale=MatchLowercase}
  \defaultfontfeatures[\rmfamily]{Ligatures=TeX,Scale=1}
\fi
\IfFileExists{upquote.sty}{\usepackage{upquote}}{}
\IfFileExists{microtype.sty}{
  \usepackage[]{microtype}
  \UseMicrotypeSet[protrusion]{basicmath} 
}{}
\makeatletter
\@ifundefined{KOMAClassName}{
  \IfFileExists{parskip.sty}{%
    \usepackage{parskip}
  }{
    \setlength{\parindent}{0pt}
    \setlength{\parskip}{6pt plus 2pt minus 1pt}}
}{
  \KOMAoptions{parskip=half}}
\makeatother
\usepackage{xcolor}
\IfFileExists{xurl.sty}{\usepackage{xurl}}{} 
\IfFileExists{bookmark.sty}{\usepackage{bookmark}}{\usepackage{hyperref}}
\hypersetup{
  colorlinks=true,
  linkcolor=Maroon,
  filecolor=Maroon,
  citecolor=Blue,
  urlcolor=Blue,
  pdfcreator={LaTeX via pandoc}}
\usepackage{longtable,booktabs}
\usepackage{etoolbox}
\makeatletter
\patchcmd\longtable{\par}{\if@noskipsec\mbox{}\fi\par}{}{}
\makeatother
\IfFileExists{footnotehyper.sty}{\usepackage{footnotehyper}}{\usepackage{footnote}}
\makesavenoteenv{longtable}
\usepackage{graphicx}
\makeatletter
\def\maxwidth{\ifdim\Gin@nat@width>\linewidth\linewidth\else\Gin@nat@width\fi}
\def\maxheight{\ifdim\Gin@nat@height>\textheight\textheight\else\Gin@nat@height\fi}
\makeatother
\setkeys{Gin}{width=\maxwidth,height=\maxheight,keepaspectratio}
\makeatletter
\def\fps@figure{htbp}
\makeatother
\providecommand{\tightlist}{%
  \setlength{\itemsep}{0pt}\setlength{\parskip}{0pt}}
\usepackage[capitalize]{cleveref}
\usepackage{CJKutf8}

\newcommand{\mean}[1]{\ensuremath{\left< #1 \right>}}
\newcommand{\Dnu}{\ensuremath{\Delta\nu}}
\newcommand{\numax}{\ensuremath{{\nu_\text{max}}}}
\newcommand{\amlt}{\ensuremath{{\alpha_\text{MLT}}}}

\newcommand{\chinesename}{{\begin{CJK}{UTF8}{gbsn}(王加冕)\end{CJK}}}

\newcommand{\annotate}[2]{\begin{tikzpicture}
    \node[anchor=south west,inner sep=0,align=center] (image) at (0,0) {
    #1
    };
    \begin{scope}[x={(image.south east)},y={(image.north west)}]
    #2
    \end{scope}
\end{tikzpicture}}
\usepackage{mathptmx,txfonts,tikz}
\ifluatex
  \usepackage{selnolig}  
\fi
\usepackage[]{natbib}
\date{\today}

\begin{document}

\shorttitle{Surface Correction Systematics}
\title{Differential Modelling Systematics across the HR Diagram from Asteroseismic Surface Corrections}
\correspondingauthor{Joel Ong}
\email{joel.ong@yale.edu}
\author[0000-0001-7664-648X]{J. M. Joel Ong \chinesename}
\affiliation{Department of Astronomy, Yale University, 52 Hillhouse Ave., New Haven, CT 06511, USA}
 \author[0000-0002-6163-3472]{Sarbani Basu}
\affiliation{Department of Astronomy, Yale University, 52 Hillhouse Ave., New Haven, CT 06511, USA}
 \author[0000-0002-6703-5406]{Jean M. McKeever}
\affiliation{Department of Astronomy, Yale University, 52 Hillhouse Ave., New Haven, CT 06511, USA}
\received{August 25, 2020}
\revised{November 3, 2020}
\accepted{November 3, 2020}
\shortauthors{Ong, Basu \& McKeever}
\def\subsectionautorefname{Section}
\def\subsubsectionautorefname{Section}


\begin{abstract}
Localised modelling error in the near-surface layers of evolutionary stellar models causes the frequencies of their normal modes of oscillation to differ from those of actual stars with matching interior structures. These frequency differences are referred to as the asteroseismic surface term. Global stellar properties estimated via detailed constraints on individual mode frequencies have previously been shown to be robust with respect to different parameterisations of this surface term. It has also been suggested that this may be true of a broader class of nonparametric treatments. We examine systematic differences in inferred stellar properties with respect to different surface-term treatments, both for a statistically large sample of main-sequence stars, as well as for a sample of red giants, for which no such characterisation has previously been done. For main-sequence stars, we demonstrate that while masses and radii, and hence ages, are indeed robust to the choice of surface term, the inferred initial helium abundance $Y_0$ is sensitive to the choice of surface correction. This implies that helium-abundance estimates returned from detailed asteroseismology are methodology-dependent. On the other hand, for our red giant sample, nonparametric surface corrections return dramatically different inferred stellar properties than parametric ones. The nature of these differences suggests that such nonparametric methods should be preferred for evolved stars; this should be verified on a larger sample.
\end{abstract}

\defcitealias{ball_correction_2014}{BG14}
\defcitealias{grevesse_standard_1998}{GS98}
\newcommand\bg{{\citetalias{ball_correction_2014}}}
\newcommand\gs{{\citetalias{grevesse_standard_1998}}}
\newcommand\mesa{{\textsc{mesa}}}
\newcommand\gyre{{\textsc{gyre}}}

\hypertarget{introduction}{%
\section{Introduction}\label{introduction}}

Asteroseismology, which has recently become popular owing to the
proliferation of high-cadence space photometry, permits precise
inference of fundamental properties of field stars, such as their masses
and radii. This is in contrast to traditional measurement techniques,
which either apply only to special dynamical configurations, such as
binary systems, or else operate on stellar populations rather than
individual stars. Such inference is predicated on our ability to
construct evolutionary models with similar enough structures to the
actual stars that their normal modes of oscillation are also similar to
the observed normal modes. On the other hand, stellar surfaces are known
to be significantly more complicated than we can adequately describe
with simple one-dimensional models. In particular, one-dimensional
computational treatments of surface magnetic fields and convection
cannot be easily reconciled with observational constraints.
Consequently, evolutionary models with theoretically identical global
properties and interior structures to the observed stars will
necessarily yield normal modes that differ slightly from to those that
would be measured observationally, owing to modelling error localised to
the stellar surface. These frequency differences are generically
referred to as the ``surface term'', and are predominantly a function of
frequency alone; in the Sun, they are known to increase in magnitude
with frequency, afflicting high-frequency modes more than low-frequency
ones
\citep{brown_solar_1984, jcd_overview_1988, dziembowski_solar_1990}.

\newcommand\yy{\left(\nu_\text{model} \over \nu_\text{ac}\right)}

Ideally, these issues may be ameliorated by explicit hydrodynamical
treatments of the stellar surface
\citep[e.g.][]{rosenthal_convective_1999, sonoi_surface_2015, mosumgaard_coupling_2020},
or through detailed analysis of the pulsation eigenfunctions
\citep[e.g.][]{roxburgh_surface_2015}. However, these prove in many
cases to be more computationally expensive than feasible for large-scale
modelling. Instead, corrections computed solely from the mode
frequencies are much more commonly used. Presently, many approaches to
detailed stellar modelling rely on a two-term surface correction
proposed in \citet[hereafter \bg]{ball_correction_2014}, of the form
\begin{equation}
    \nu_\text{corr} = \nu_\text{model} + {\nu_\text{ac} \over I_\nu}\left(a_{-1} \yy^{-1} + a_3 \yy^3\right), \label{eq:bg14}
\end{equation} where \(I_\nu\) is the normalisation-independent,
dimensionless mode inertia, and the acoustic cutoff frequency
\(\nu_\text{ac}\) may be replaced with \numax~up to rescaling of the
coefficients \(a_{-1}\) and \(a_3\). These are left to vary freely, and
are fitted as additional free parameters in determining a best-fitting
model. This construction is motivated by the functional analysis of
\citet{gough_comments_1990}. While the original analysis applied to the
Sun (and by extension Sun-like stars), \citet{ball_surface_2017}
establish the viability of using the \bg~correction for evolved stars as
well.

Such a parameterisation of the surface term is not unique, and naturally
there exist competing parameterisations
\citep[e.g.][]{kjeldsen_correcting_2008, sonoi_surface_2015}. However,
the two-term parameterisation of \bg~is more physically motivated than
most others, which are instead largely phenomenological in nature, and
has been shown empirically to be quite robust. Comparing the performance
of different parametric surface corrections (including that of \bg)
across the HR diagram in a synthetic test, \citet{schmitt_modeling_2015}
concluded that the two-term correction of \bg~yields the most robust
agreement with synthetic surface terms generated by perturbing
theoretical models. Similarly, \citet{nsamba_asteroseismic_2018} and
\citet{compton_surface_2018} found the two-term correction of \bg~to be
the most likely, in a Bayesian sense, when applied to the Kepler LEGACY
sample. \citet{ball_surface_2017} arrived at a similar conclusion for a
small sample of red giants, as did \citet{ball_surface_2018}. Likewise,
\citet{jorgensen_investigating_2020}, who compared seismic masses from
various surface corrections with dynamical masses from eclipsing
binaries, again concluded that the two-term correction is more robust
than other competing parameterisations.

While these works have looked only at various explicit parameterisations
of the surface term, there also exist several nonparametric diagnostics.
For example, in lieu of the frequencies themselves,
\citet{roxburgh_ratio_2003} propose the use of frequency separation
ratios as a surface-insensitive set of observational constraints
\citep[see also][]{otifloranes_use_2005}, although the construction of
these quantities introduces correlated errors
\citep{roxburgh_overfitting_2018}. \citet{roxburgh_asteroseismic_2016}
also proposes an alternative, less commonly employed phase-matching
algorithm. Regarding the use of these various diagnostics and
corrections in inferring fundamental stellar properties like masses,
radii and compositions, \citet{nsamba_asteroseismic_2018} and
\citet{basu_robustness_2018} demonstrate that such parameter inferences
are insensitive to these methodological choices, at least on the main
sequence. However, they leave open the question as to whether such
robustness continues to obtain for more evolved stars.

Fundamental properties inferred from stellar modelling are known to
suffer from various systematic errors owing to both degeneracies in the
observational constraints, and uncertainties in the underlying physics.
For example, in the absence of a luminosity constraint, posterior
distributions returned from grid-based modelling are known to suffer
strong correlations between the inferred posterior masses and initial
helium abundances, which in turn leads to larger age uncertainties than
would be expected from simply propagating the mass uncertainties.
Likewise, mixing-length treatments of convective efficiency, which are
commonly used in one-dimensional modelling, yield an additional model
parameter \amlt~that is likewise strongly correlated with the inferred
masses and radii. In many stellar modelling applications, these
parameters are either held fixed (e.g.~with a solar-calibrated mixing
length) or constrained to reduce the dimensionality of the parameter
space \citetext{\citealp[e.g.~with a metallicity-mixing length
relation,][]{viani_investigating_2018}; \citealp[or a metallicity-helium
enrichment relation,
e.g.][]{moedas_asteroseismic_2020}; \citealp{nsamba_asteroseismic_2020}}.

The imposition of seismic constraints may reduce the overall
uncertainties of the inferred quantities, but does not a priori
eliminate such degeneracy between model parameters. Rather, seismic
constraints may induce new systematic correlations. For example, the
large separation \(\Delta\nu\) is (approximately) a direct constraint on
the mean density, and so yields strong degeneracy between the inferred
masses and radii. This degeneracy is subtly modified by the surface
term; e.g.~the use of the \bg~correction is known to modify
\Dnu~slightly \citep{viani_investigating_2018}. The effects of other
surface corrections are less well-studied. In this paper, we investigate
the extent and nature of modifications to these systematics introduced
by different choices of surface correction.

\hypertarget{modelling-procedure}{%
\section{Modelling procedure}\label{modelling-procedure}}

\label{sec:modelling}

The inference of stellar properties from global seismic quantities is
known to be affected by a very large number of model parameters,
including the initial helium abundance, mixing length parameter, and
other choices of model physics --- e.g.~core or envelope convective
overshoot, element diffusion, or the abundance mixture \citep[see
e.g.][]{lebreton_accurate_2014, reese_hh_2016, silvaaguirre_standing_2017, nsamba_asteroseismic_2018}.
In this work, we focus largely on the first two parameters. In order to
consistently isolate surface-correction-induced systematic effects from
those resulting from the underlying stellar models, we adopt a
grid-based modelling and Bayesian inference procedure, where likelihoods
for all stars in a given sample are inferred from the same set of
stellar models. This is as opposed to a sampling scheme dictated by an
optimisation-based modelling procedure \citep[e.g.~several of the
methods in][]{silvaaguirre_standing_2017} which would have yielded a
different set of models for each target, or interpolation-based grid
modelling \citep[e.g.][]{rendle_aims_2019}, which is significantly more
computationally expensive.

\hypertarget{cost-function}{%
\subsection{Cost function}\label{cost-function}}

\label{sec:costs}

For each star, we should, in principle, compute a log-likelihood
function for every model in the grid out of contributions from
spectroscopic and seismic inputs, as \begin{equation}
    \chi^2 = {1 \over 2} \left(\chi^2_{T_\text{eff}} + \chi^2_{\text{[Fe/H]}} + \chi^2_\text{seis}\right)
\end{equation} This choice of normalisation (in particular, using the
reduced instead of total \(\chi^2\)) was made for consistency with the
procedure of \citet{mckeever_helium_2019}, whose sample of red giants we
also use in this work. For the seismic cost term, we choose from one of
the following:

\begin{itemize}
\tightlist
\item
  A cost term from the (parametric) surface correction of
  \citet{ball_correction_2014};
\item
  A cost term from the (nonparametric) phase matching algorithm of
  \citet{roxburgh_asteroseismic_2016}; or
\item
  A (nonparametric) cost term constructed from the dipole and quadrupole
  separation ratios \citep[e.g.][]{roxburgh_ratio_2003}.
\end{itemize}

\begin{figure}[htbp]
    \centering
    \annotate{\includegraphics[width=.45\textwidth, trim=.25cm .25cm .25cm .15cm,clip]{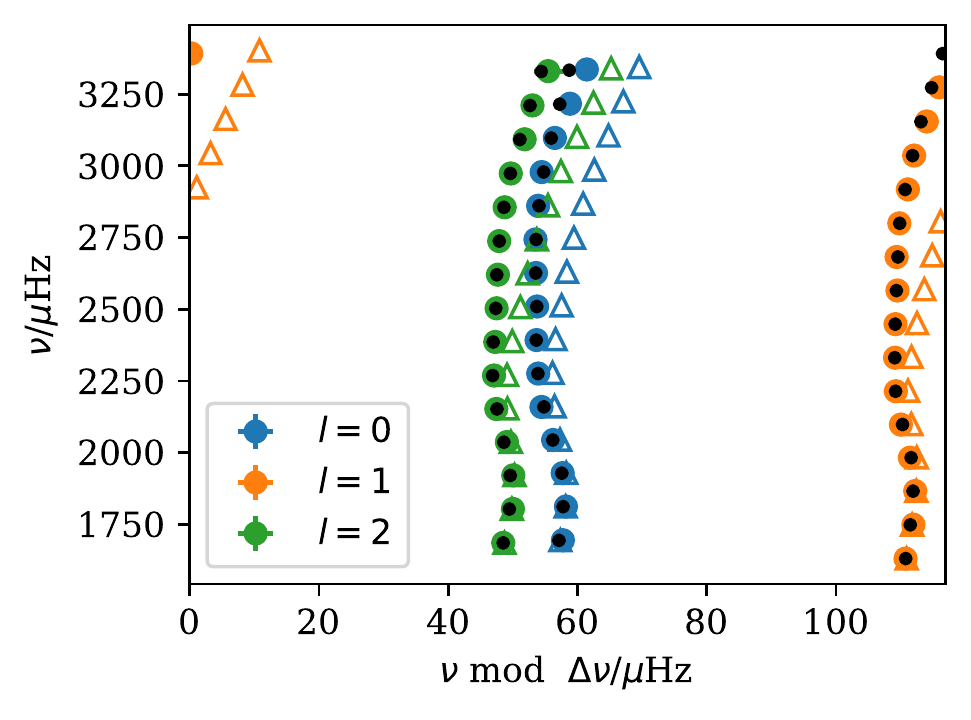}}{\node at (.9, .22){\textbf{(a)}};}
    \annotate{\includegraphics[width=.45\textwidth, trim=.25cm .25cm .25cm .15cm,clip]{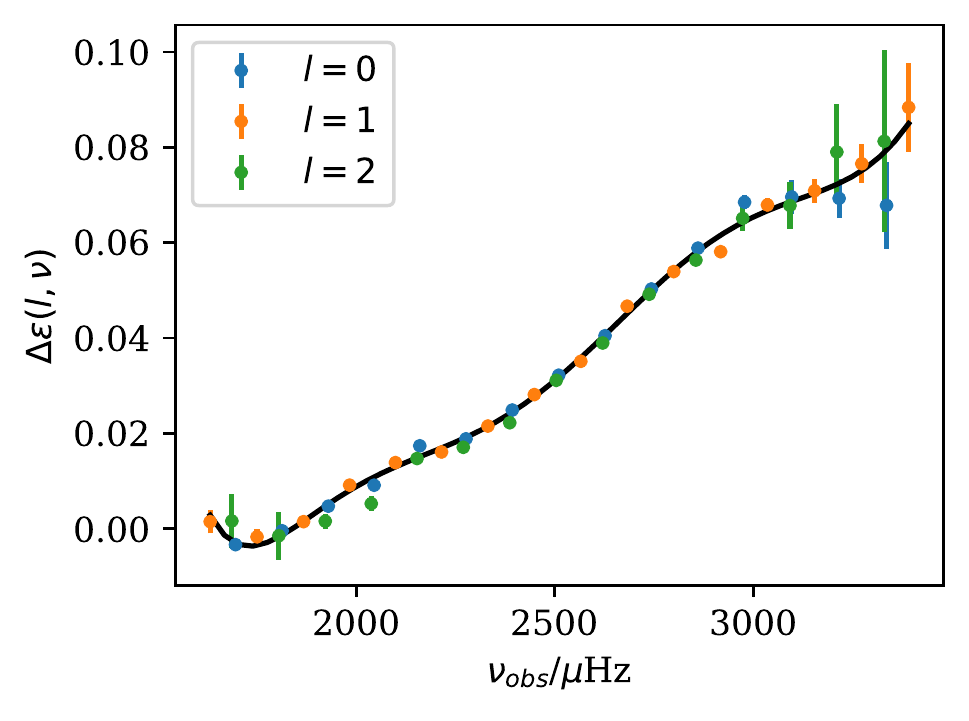}}{\node at (.95, .22){\textbf{(b)}};}
    \annotate{\includegraphics[width=.45\textwidth, trim=.25cm .25cm .25cm .15cm,clip]{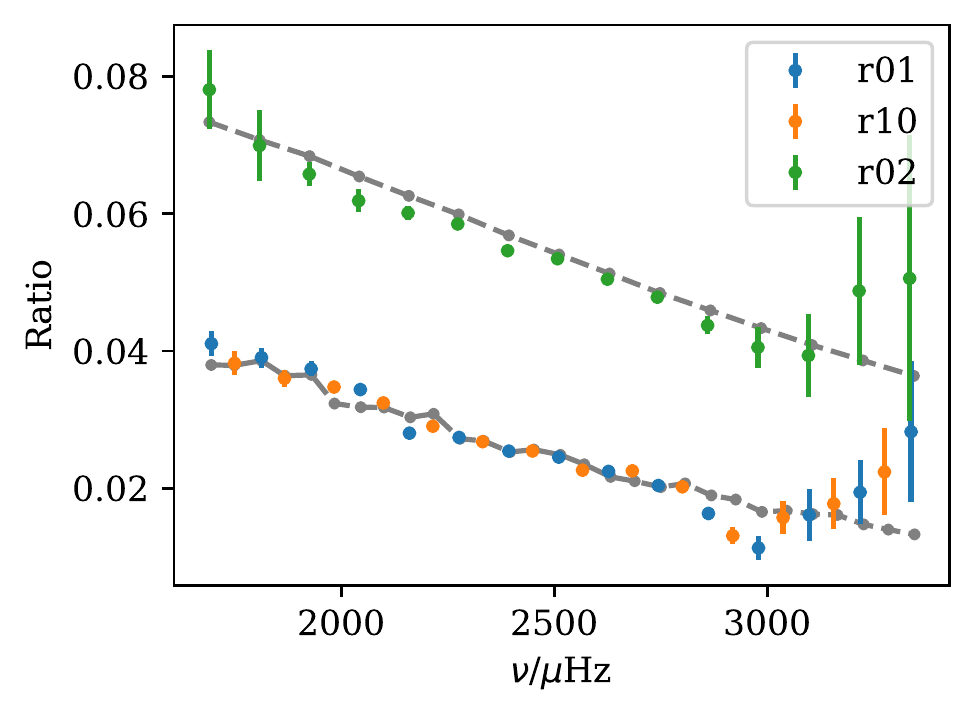}}{\node at (.22, .22){\textbf{(c)}};}
    \caption{Various surface corrections applied to 16 Cyg B, one of the stars in the LEGACY sample. \textbf{(a)}: Echelle diagram showing observed frequencies with filled circles, uncorrected model frequencies from a good-fitting model with open triangles, and corrected frequencies (per \cref{eq:bg14}) with black dots. \textbf{(b)}: Degree-wise phase differences and fitted Chebyshev polynomial for the same model; the various phase-difference functions have collapsed close to a single function of frequency. \textbf{(c)}: Various frequency ratios. Model quantities are shown with grey dots, joined with dashed lines.}
    \label{fig:surface}
\end{figure}

However, additional regularisation is necessary when employing the
phase-matching or separation-ratio methods, which do not discriminate
between stars of different large separations
\citep{roxburgh_asteroseismic_2016}. Now, from helioseismology it is
known that, at least for the Sun, the surface term decreases with
frequency, leaving the lowest-frequency modes least affected
\citep{dziembowski_solar_1990}. This is typically also already the case
for our best-fitting models using the \bg~surface correction (see
e.g.~\cref{fig:surface}a). Therefore, assuming this to be true for the
surface term in general, we include a regularisation term
\(\chi^2_\text{reg}\), defined to be the reduced chi-squared associated
with the \(N_\text{reg}\) lowest-frequency uncorrected modes:
\begin{equation}
    \chi^2_\text{reg} = {1 \over N_\text{reg} - 1}\sum_j^{N_\text{reg}}\left(\nu_j - \nu_\text{mod,j} \over \sigma_{\nu,j}\right)^2.
\end{equation} For our main-sequence sample, we choose
\(N_\text{reg} = 6\), while for our red giants, which have fewer
detected modes on average, we choose \(N_\text{reg} = 2\). This was also
done in \citet{basu_robustness_2018}, who used \(N_\text{reg}=2\) for
all stars.

With the inclusion of this term, our cost function then takes the form
\begin{equation}
    \chi^2 = {1 \over 3} \left(\chi^2_{T_\text{eff}} + \chi^2_{\text{[Fe/H]}} + \chi^2_\text{reg} + \chi^2_\text{seis}\right).\label{eq:cost}
\end{equation} We summarise each of the seismic constraints below. To
illustrate these procedures, we show the corresponding quantities
associated with one good-fitting stellar model for one of the stars in
our sample (KIC 12069449, 16 Cyg B) in \cref{fig:surface}.

\hypertarget{parametric-surface-correction}{%
\subsubsection{\texorpdfstring{Parametric surface correction
\label{sec:bgcost}}{Parametric surface correction }}\label{parametric-surface-correction}}

We adopt the \bg~surface correction, \cref{eq:bg14}, as our reference
parametric surface correction, as it does not require calibration
against a reference star, and, as described previously, has been shown
to be both more robust and more physically motivated than most other
extant parametric surface corrections. In particular, such frequency
differences were derived in \citet{gough_comments_1990} from
perturbation theory. This does not guarantee the converse --- in
general, structural differences confined to the near-surface layers may
not necessarily result in frequency shifts of these forms. Nonetheless,
this is assumed as an ansatz underlying the method.

We construct a reduced-\(\chi^2\) function in the usual manner, as
\begin{equation}
    \chi^2_\text{BG} = {1 \over N - 3} \sum_i^N \left(\nu_\text{obs,i} - \nu_\text{corr,i} \over \sigma_{\nu_i}\right)^2,\label{eq:bgcost}
\end{equation} where the corrected frequencies are as specified in
\cref{eq:bg14}. The parameters \(a_{-1}\) and \(a_3\) for each stellar
model were chosen to minimize this quantity as computed with respect to
only radial modes, while the cost functions used in the grid search were
computed for each model with respect to those same parameter values, but
with all modes included.

\hypertarget{nonparametric-surface-correction-phase-matching}{%
\subsubsection{Nonparametric surface correction: phase
matching}\label{nonparametric-surface-correction-phase-matching}}

\citet{roxburgh_ratio_2003} constructed a partial-wave eigenvalue
equation for stellar p-modes, of the form \begin{equation}
    \omega T - \alpha_l(\omega) + \delta_l(\omega) \equiv \omega T - \pi \epsilon_l(\omega) = \left(n_p + {l \over 2}\right) \pi,\label{eq:partial}
\end{equation} with eigenvalues recovered for integer \(n_p\). The
quantities \(\alpha_l\) and \(\delta_l\) are the partial-wave phase
functions of degree \(l\) associated with the outer and inner boundary,
respectively, which may be separately computed by integrating a
nonlinear initial value problem from their respective boundaries to a
common matching point in the interior.

As per \citet{roxburgh_asteroseismic_2016}, two stars which differ from
each other only in the structure of their outer layers will have
identical inner phase functions, and the differences of their outer
phase functions \(\alpha_l\), and therefore total phase functions
\(\epsilon_l\), will be independent of degree (as in
\cref{fig:surface}b). Again, this does not guarantee the converse: a
degree-independent phase-difference function does not in itself
guarantee that model differences are localised to the surface. Again,
this is nonetheless assumed as part of the method.

Phase functions for modes of each degree are constructed for each model
in the grid as \begin{equation}
    \epsilon_l(\nu_{nl}) = {\nu_{nl}\over\Delta\nu} - n - {l \over 2},
\end{equation} treated as samples of some function of frequency. These
are interpolated with a cubic spline, and evaluated at the observed
frequencies. For each model we then compute the cost function as the
minimum value of \begin{equation}
    \chi^2_\epsilon = {1 \over N - M - 1} \sum_l \sum_i^{N_l}\left(\epsilon_l^\text{model}(\nu_{n_il}) - \epsilon_{n_il}^\text{obs} - F_M(\nu_{n_il}) \over \sigma_{\epsilon,n_il}\right)^2,
\end{equation} where \(F_M\) is is a degree-independent function of
frequency with at most \(M < N_0\) parameters. Common choices for such
functions include B-splines, or linear combinations of various families
of orthogonal polynomials; for our modelling we have opted to follow
\citet{roxburgh_asteroseismic_2016} in using a linear combination of
Chebyshev polynomials.

\hypertarget{nonparametric-surface-correction-separation-ratios}{%
\subsubsection{Nonparametric surface correction: separation
ratios}\label{nonparametric-surface-correction-separation-ratios}}

\label{sec:ratios}

Using \cref{eq:partial}, \citet{roxburgh_ratio_2003} further
demonstrated that various frequency separation ratios can be related to
only the inner partial-wave phase functions of different degrees,
evaluated at the same fiducial frequencies. For instance, to leading
order, \begin{equation}
    r_{02}(n) = {\nu_{n, 0} - \nu_{n-1, 2}\over \nu_{n, 1} - \nu_{n-1, 1}} \sim {1 \over \pi} \left(\delta_2(\nu_{n,02}) - \delta_1(\nu_{n,02})\right),
\end{equation} where the \(\delta_l\) are the inner phase functions of
\cref{eq:partial}, evaluated at the fiducial frequency
\(\nu_{n,02} = (\nu_{n, 0} - \nu_{n-1, 2})/2\). Similar constructions
exist for other separation ratios like \(r_{01}\).

To evaluate the corresponding cost function, we compute these separation
ratios separately for the model and observed frequencies. These are to
be interpreted as two different functions of frequency
\citep{roxburgh_scaled_2015}. For each model, we interpolate its
separation ratio function to the observed fiducial frequencies with a
cubic spline (as in \cref{fig:surface}c), and evaluate its cost as,
e.g., \begin{equation}
    \chi^2_{r_{02}} = {1 \over N_{r_{02}} - 1} \sum_j^{N_{r_{02}}}\left( r_{02,\text{model}}(\nu_{j,02}) - r_{02, \text{obs}, j}\over \sigma_{r_{02},j}\right)^2.
\end{equation} We use only \(r_{01}\) and \(r_{02}\) in this manner
\citep[as \(r_{01}\) and \(r_{10}\) contain the same information with
strongly correlated errors; see][]{roxburgh_overfitting_2018}, combining
them as \(\chi^2_\text{ratio} = (\chi^2_{r_{01}} + \chi^2_{r_{02}})/2\).
For the dipole separation ratios in particular, we use a three-point
rather than five-point construction \citep[see][]{basu_robustness_2018}
to further reduce correlated errors.

\hypertarget{monte-carlo-recovery-of-posterior-distribution}{%
\subsection{Monte-Carlo recovery of posterior
distribution}\label{monte-carlo-recovery-of-posterior-distribution}}

\begin{figure*}
\centering
\includegraphics{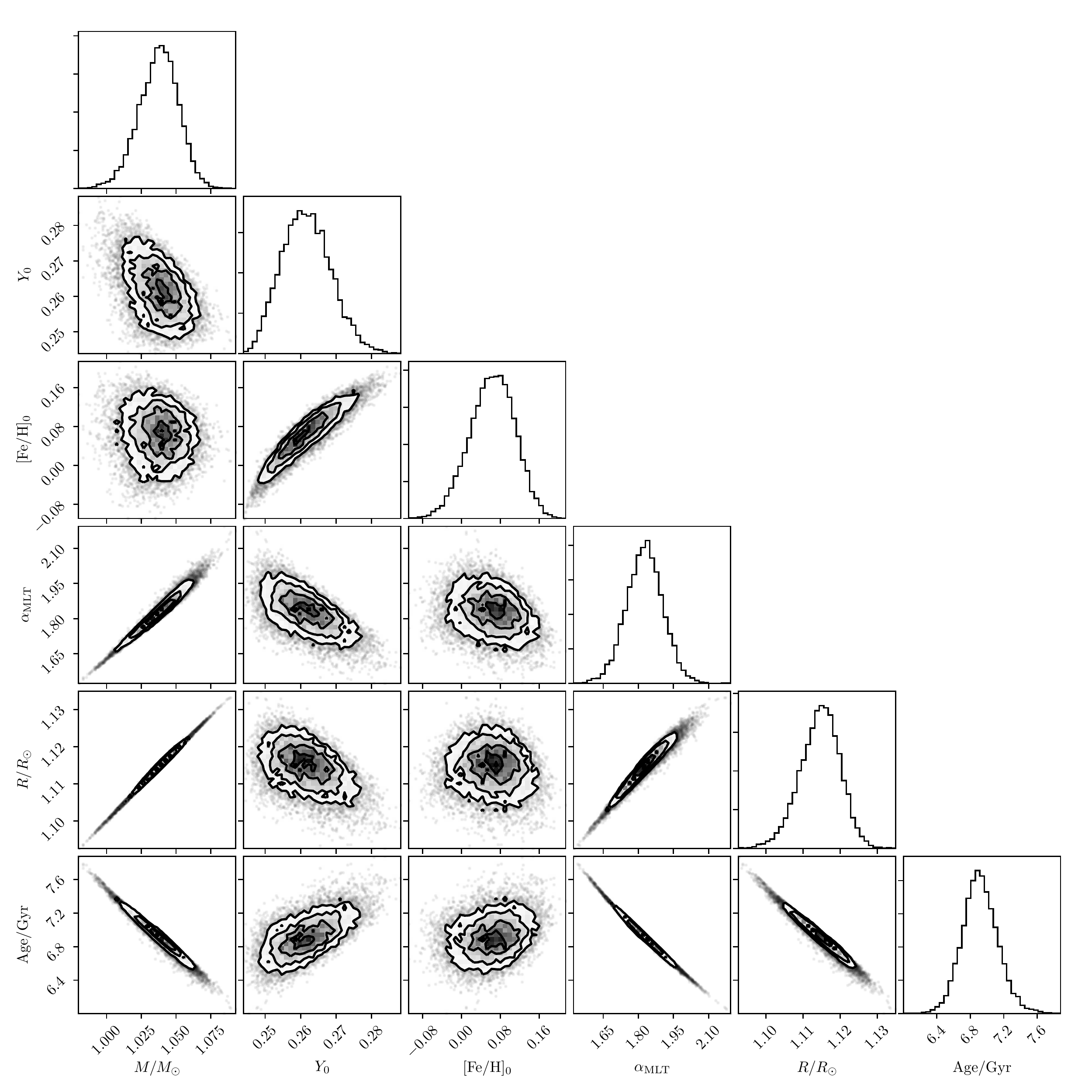}
\caption{Corner plot for 16 Cyg B, showing joint and marginal posterior distributions for various stellar fundamental parameters, using the \bg\ surface correction; shown here are the mass, initial helium abundance, initial metallicity, mixing-length parameter, radius, and age.\label{fig:corner}}
\end{figure*}

For each star, given a set of spectroscopic constraints and a grid of
stellar models, we treat the cost functions computed in the preceding
section for each model as a log-likelihood function over the grid:
\(\mathcal{L}_i \propto \exp\left[-\chi^2_i/2\right]\). Given a prior
distribution over the parameter space of the grid, this suffices to
define a formal posterior distribution \begin{equation}
    p_i \propto \mathcal{L}_i w_i,\label{eq:weights}
\end{equation} where \(w_i\) is inversely proportional to the assumed
prior distribution. We may then find weighted averages of, e.g., the
stellar mass, by averaging over the grid as e.g.~\begin{equation}
    M \sim \sum_i p_i M_i. \label{eq:mean}
\end{equation} In principle, this would also permit us to recover higher
moments, such as the posterior covariances.

However, depending on the sampling method chosen for the grid, the
resulting posterior distribution functions may not be well-behaved;
e.g.~the model grid may only sample a few discrete values of
metallicity, helium abundance, or mixing length. Moreover, strictly
speaking the form of \cref{eq:cost}, with its various terms normalised
differently, cannot be rigorously interpreted as a true log-likelihood
function in the usual sense.

To circumvent these difficulties, we estimate the posterior
distributions of these parameters using a Monte-Carlo scheme. In
particular, we repeat the computation above a large number of times,
each time perturbing each of the spectroscopic values from the nominal
data by one realisation of independent and identically distributed
Gaussian random quantities, with zero mean and with variances given by
the nominal measurement errors \citep[as in
e.g.][]{basu_determination_2010}. We collect the weighted averages
associated with every one of these realisations. Our final estimate of
the posterior distribution is then taken to be the distribution of the
weighted means returned from all of these realisations. We use \(10^4\)
realisations for this bootstrapping procedure, which appears to be a
good compromise between being large enough to converge to the equlibrium
distribution, while also not taking too long to compute for each star.
We illustrate the result of this procedure in \cref{fig:corner}.

\hypertarget{main-sequence-stars-the-legacy-sample}{%
\section{Main sequence stars: the LEGACY
sample}\label{main-sequence-stars-the-legacy-sample}}

For studying main-sequence stars, we use the LEGACY sample
\citep{lund_legacy_2017}, consisting of 66 targets observed with
\emph{Kepler} in short (60 second) cadence. For our modelling, we used
spectroscopic constraints derived from high-resolution spectroscopy
under the Kepler Follow-up Program \citep[KFOP,][]{furlan_kepler_2018},
which was available for 59 out of the 66 stars in the sample. For the
remainder, we used the nominal values from the original LEGACY survey
\citep{silvaaguirre_standing_2017}. We also use the same detailed mode
frequencies as \citet{silvaaguirre_standing_2017}.

\hypertarget{model-grid}{%
\subsection{Model grid}\label{model-grid}}

For the purposes of stellar modelling, we construct a Sobol sequence of
\(2^{14} -1\) elements across a parameter space consisting of the
stellar mass \(M/M_\Sun \in [0.7, 1.7]\), initial helium abundance
\(Y_0 \in [0.23, 0.37]\), initial metallicity
\([\text{Fe/H}]_0 \in [-1.5, +0.25]\), and the mixing length parameter,
spaced logarithmically as
\(\log_{10} \amlt/\alpha_{\text{MLT},\Sun} \in [-0.2, +0.2]\). For each
element in the Sobol sequence we constructed a series of stellar models
with \mesa~r12115, using an Eddington gray atmospheric boundary
condition, the mixing-length prescription of
\citet{cox_principles_1968}, the relative heavy-element abundances of
\citet[][\gs]{grevesse_standard_1998}, with the diffusion and settling
of helium and heavy elements as in \citet{burgers_flow_1969}, and
without any convective overshoot, either in the envelope or core.
Performing a solar calibration with these choices of physics yielded a
solar-calibrated mixing length of \(1.82\) and an initial helium
abundance of \(0.272\). A second copy of this Sobol sequence was also
constructed with identical values, except for higher metallicities in
the range \([+0.25, +2.0]\). Only points with
\([\text{Fe/H}]_0 < +0.75\) were used from this second sequence.
Evolutionary tracks were terminated either at a maximum age of 20 Gyr,
or 0.5 Gyr after core hydrogen exhaustion, whichever was earlier. Our
use of the \gs~solar elemental abundances is motivated by consistency
with the community consensus values in the original LEGACY modelling
effort \citep{silvaaguirre_standing_2017}, which were also derived with
respect to \gs~abundances (except in one case where
\citet{grevesse_atomic_1993} abundances were used). These abundances are
known to yield better consistency with helioseismic
\citep{antia_determining_2006, basu_helioseismology_2008, basu_revisiting_2013}
and \(pp\)-neutrino \citep{vinyoles_new_2017} constraints than do more
recent alternatives, such as the abundances of \citet{agss09}.

To reduce the computational time required to compute normal mode
frequencies, we did the pulsation calculations in a two-step process.
Each evolutionary track was run twice: once under default adaptive
timestep controls, and a second time with a maximum timestep of 0.03
Gyr. For the first run, the frequencies and inertiae of normal modes
within \(\pm 12 \Dnu\) of \(\numax\) were computed for \(l = 0, 1, 2\)
using \gyre~v5.2. The frequencies of the second set of models were then
derived by interpolating those of the first set in dimensionless units,
as described in \citet{rendle_aims_2019}. The mode inertiae were also
interpolated logarithmically. In all, the second run yielded a total of
about 5 million models.

For a discretely-sampled grid, it is common to check how the inferred
parameters change for different values of some model parameters, which
are otherwise held fixed. For example, historically \amlt~could not be
observationally constrained, and most isochrone calculations are
performed with a fixed value of \amlt. On the other hand, seismic radii
are known to depend on \amlt, and so we might wish to compare the radii
obtained from constraining \amlt~to different values. This is typically
done by comparing the properties inferred with respect to only models
with \(x_i = x_{i,0}\), where \(x_i\) is a coordinate of the grid. Since
doing this with our grid would discard almost all of the Sobol sequence,
we cannot apply this procedure directly. Instead, we recognise that such
a restriction to the models selected from the grid amounts to
constructing means and uncertainties associated with conditional
probability distributions. This can otherwise also be effected by
applying additional weights to the grid. Selecting a fixed value of
\(x_i = x_{i,0}\) with an ordinary grid approximates the analytic
expression \begin{equation}
    p\left(x | x_i = x_{i, 0}\right) \propto p(x) \delta(x_i - x_{i,0})\label{eq:conditional},
\end{equation} so that \begin{equation}
    \mean{F}_{x_i=x_{i,0}} = \int \mathrm d^n x\ F(x) p(x | x_i=x_{i,0}).
\end{equation} For our Sobol-sequence set of evolutionary tracks, we
approximate the \(\delta\) distribution by using Gaussian weights in
\cref{eq:weights} such that \begin{equation}
    w_j = \exp \left[-{(x_{i,j} - x_{i,0})^2 \over 2 \sigma^2}\right]. \label{eq:conditional2}
\end{equation} In most of our applications below, the width of the
Gaussian is chosen such that the \(\pm1\sigma\) region contains as many
sample points as would a slice of a uniformly-sampled discrete grid with
the same number of grid points, which works out to about 10\% of the
Sobol samples in our case.

\begin{figure}
\centering
\includegraphics{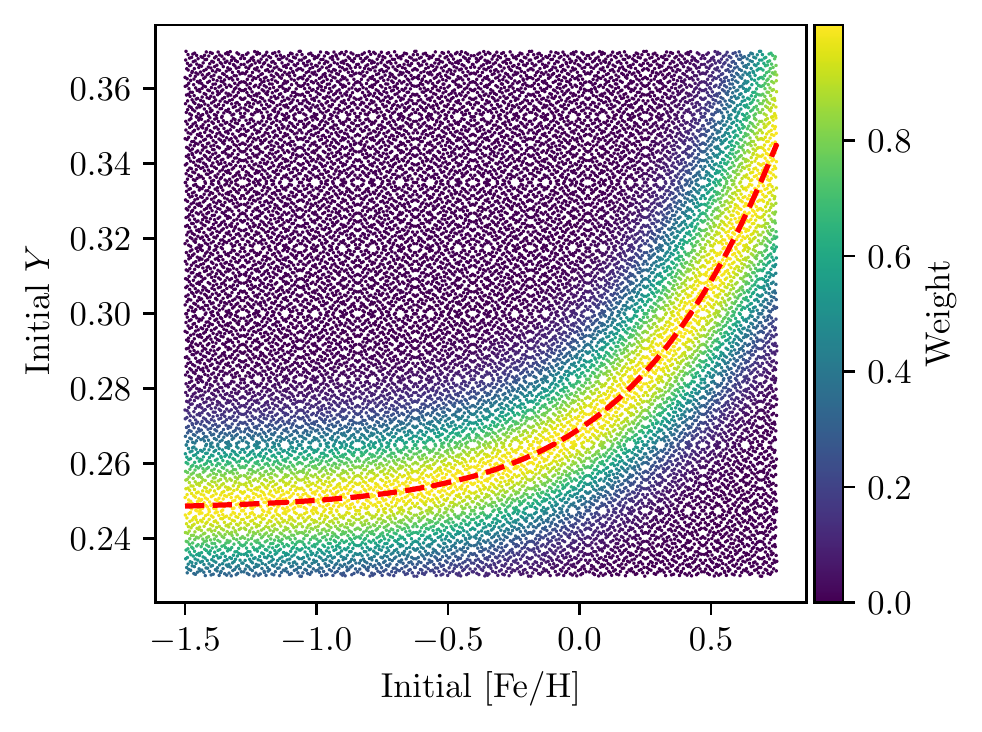}
\caption{Weights assigned to Sobol sequence points in our model grid for
computing conditional distributions associated with a solar-like
enrichment law (i.e.~fixed \(\mathrm{d}Y/\mathrm{d}Z\) relation, shown
in these coordinates with a red dashed line). \label{fig:dYdZ}}
\end{figure}

Parameterising the selection of restricted sets of stellar models as
conditional probabilities in this manner also permits more flexible
constraints on model selection. For example, since the initial helium
abundance is difficult to constrain observationally, it is often
determined a priori via a linear enrichment law of the form
\begin{equation}
    Y_0 = Z {\mathrm d Y \over \mathrm d Z} + Y_i,\label{eq:enrichment}
\end{equation} where \(Y_i\) is set to the primordial value of 0.248,
and with the gradient chosen, for example, to yield a star of solar
metallicity at the solar age when \(Y_0\) is at the solar-calibrated
value. Comparing the inferred stellar properties returned under such a
model selection constraint to those returned where \(Y_0\) is permitted
to vary freely typically necessitates the generation of two different
model grids
\citep[e.g.][]{moedas_asteroseismic_2020, nsamba_asteroseismic_2020}. By
contrast, we may instead constrain the initial helium abundance and
metallicity to to lie on a hypersurface of constant \(M\) and \(\alpha\)
(whose projection to the helium-metallicity plane we show in
\cref{fig:dYdZ}), corresponding to the image of \cref{eq:enrichment} in
the natural coordinates of the Sobol sequence. The distance function
used for the Gaussian weights in \cref{eq:conditional2} is replaced by
the minimum Euclidean distance from each grid point to this hypersurface
with respect to these coordinates. The resulting weights are also shown
in \cref{fig:dYdZ}.

\hypertarget{differences-in-inferred-parameters}{%
\subsection{Differences in inferred
parameters}\label{differences-in-inferred-parameters}}

We wish to determine if different choices of surface correction impart
systematic differences to the resulting estimates of stellar masses,
radii, ages, etc. While earlier studies have looked at this in detail on
a somewhat ad hoc star-by-star basis
\citep[e.g.][]{nsamba_asteroseismic_2018, compton_surface_2018, basu_robustness_2018},
we opt instead to take a more conservative statistical approach.

Let us consider pairwise differences between the posterior mean masses
returned from grid-based modelling with respect to two different surface
corrections (e.g.~that of \bg~vs.~the phase matching method). For each
star indexed by \(j\), we define a normalised difference in e.g.~the
stellar mass, as \begin{equation}
    z_{M, j, \text{BG vs. }\epsilon} = {M_{j, \epsilon} - M_{j, \text{BG}} \over \sqrt{\sigma_{j,M,\epsilon}^2 + \sigma_{j,M,\text{BG}}^2}}, \label{eq:z}
\end{equation} where \(M\) and \(\sigma_M\) represent the posterior
means and standard deviations. If the surface correction does not affect
the result of stellar modelling, then in principle this quantity
(assembled over all stars in the sample) should be distributed with zero
mean. We therefore take this (i.e.~the statement that
\(z_{M, \text{BG vs. }\epsilon}\) is distributed with zero mean) as our
null hypothesis. Our alternate hypothesis is that the population mean is
non-zero, so we perform a one-sample two-tailed \(t\)-test \citep[as
done in e.g.][]{silvaaguirre_standing_2017}. As a reminder: the
\(p\)-value returned from such a hypothesis test represents the
probability (in a frequentist sense) of making a Type I error ---
i.e.~incorrectly rejecting the null hypothesis. It is customary to
define in advance a required level of significance \(\alpha\), so as to
reject the null hypothesis only for \(p < \alpha\). For the purposes of
discussion, we adopt \(\alpha = 0.0025\), corresponding to \(3\sigma\)
significance on a two-tailed test. However, we will see that many cases
are unambiguous.

We show in \cref{tab:t1} the \(p\)-values resulting from comparing
different surface corrections, including under the effects of different
constraints on model selection from the grid (in particular, with a
restriction to solar-calibrated mixing length, or a solar-like linear
enrichment law). Each entry in the table contains two numbers: the first
for a comparison of \bg~vs.~phase-matching, and the second for
\bg~vs.~separation ratios. We do this for the stellar mass, radius, age,
and initial helium abundance.

\newcommand\subrow{{\footnotesize$\begin{array}{c}\text{\bg\ vs. }\epsilon \\ \text{\bg\ vs. } r\end{array}$}}

\begin{longtable}[]{@{}lrccc@{}}
\caption{\(t\)-test \(p\)-values for differences between surface
corrections (under the null hypothesis that the differences \(z\),
\cref{eq:z}, are distributed with a population mean of
\(0\)).\label{tab:t1}}\tabularnewline
\toprule
\begin{minipage}[b]{0.08\columnwidth}\raggedright
Quantity\strut
\end{minipage} & \begin{minipage}[b]{0.15\columnwidth}\raggedleft
\strut
\end{minipage} & \begin{minipage}[b]{0.18\columnwidth}\centering
Full Grid\strut
\end{minipage} & \begin{minipage}[b]{0.23\columnwidth}\centering
Solar Enrichment (1)\strut
\end{minipage} & \begin{minipage}[b]{0.21\columnwidth}\centering
Solar Mixing Length (2)\strut
\end{minipage}\tabularnewline
\midrule
\endfirsthead
\toprule
\begin{minipage}[b]{0.08\columnwidth}\raggedright
Quantity\strut
\end{minipage} & \begin{minipage}[b]{0.15\columnwidth}\raggedleft
\strut
\end{minipage} & \begin{minipage}[b]{0.18\columnwidth}\centering
Full Grid\strut
\end{minipage} & \begin{minipage}[b]{0.23\columnwidth}\centering
Solar Enrichment (1)\strut
\end{minipage} & \begin{minipage}[b]{0.21\columnwidth}\centering
Solar Mixing Length (2)\strut
\end{minipage}\tabularnewline
\midrule
\endhead
\begin{minipage}[t]{0.08\columnwidth}\raggedright
Mass\strut
\end{minipage} & \begin{minipage}[t]{0.15\columnwidth}\raggedleft
\subrow\strut
\end{minipage} & \begin{minipage}[t]{0.18\columnwidth}\centering
\(\begin{array}{c}0.4530 \\ 0.5835\end{array}\)\strut
\end{minipage} & \begin{minipage}[t]{0.23\columnwidth}\centering
\(\begin{array}{c}0.2783 \\ 0.2967\end{array}\)\strut
\end{minipage} & \begin{minipage}[t]{0.21\columnwidth}\centering
\(\begin{array}{c}0.5836 \\ 0.2815\end{array}\)\strut
\end{minipage}\tabularnewline
\begin{minipage}[t]{0.08\columnwidth}\raggedright
Radius\strut
\end{minipage} & \begin{minipage}[t]{0.15\columnwidth}\raggedleft
\subrow\strut
\end{minipage} & \begin{minipage}[t]{0.18\columnwidth}\centering
\(\begin{array}{c}0.7802 \\ 0.1088\end{array}\)\strut
\end{minipage} & \begin{minipage}[t]{0.23\columnwidth}\centering
\(\begin{array}{c}0.0245 \\ 0.0407\end{array}\)\strut
\end{minipage} & \begin{minipage}[t]{0.21\columnwidth}\centering
\(\begin{array}{c}0.7242 \\ 0.1428\end{array}\)\strut
\end{minipage}\tabularnewline
\begin{minipage}[t]{0.08\columnwidth}\raggedright
Age\strut
\end{minipage} & \begin{minipage}[t]{0.15\columnwidth}\raggedleft
\subrow\strut
\end{minipage} & \begin{minipage}[t]{0.18\columnwidth}\centering
\(\begin{array}{c}0.9392 \\ 0.5293\end{array}\)\strut
\end{minipage} & \begin{minipage}[t]{0.23\columnwidth}\centering
\(\begin{array}{c}0.7370 \\ 0.4890\end{array}\)\strut
\end{minipage} & \begin{minipage}[t]{0.21\columnwidth}\centering
\(\begin{array}{c}0.1862 \\ 0.1073\end{array}\)\strut
\end{minipage}\tabularnewline
\begin{minipage}[t]{0.08\columnwidth}\raggedright
\(Y_0\)\strut
\end{minipage} & \begin{minipage}[t]{0.15\columnwidth}\raggedleft
\subrow\strut
\end{minipage} & \begin{minipage}[t]{0.18\columnwidth}\centering
\(\begin{array}{c}1.82 \times 10^{-8} \\ 0.0110\end{array}\)\strut
\end{minipage} & \begin{minipage}[t]{0.23\columnwidth}\centering
\(\begin{array}{c}0.0166 \\ 0.1645\end{array}\)\strut
\end{minipage} & \begin{minipage}[t]{0.21\columnwidth}\centering
\(\begin{array}{c}6.85 \times 10^{-8} \\ 0.9593\end{array}\)\strut
\end{minipage}\tabularnewline
\bottomrule
\end{longtable}

\begin{figure}[htbp]
    \centering
    \includegraphics[width=.45\textwidth, trim=.25cm .25cm .25cm .15cm,clip]{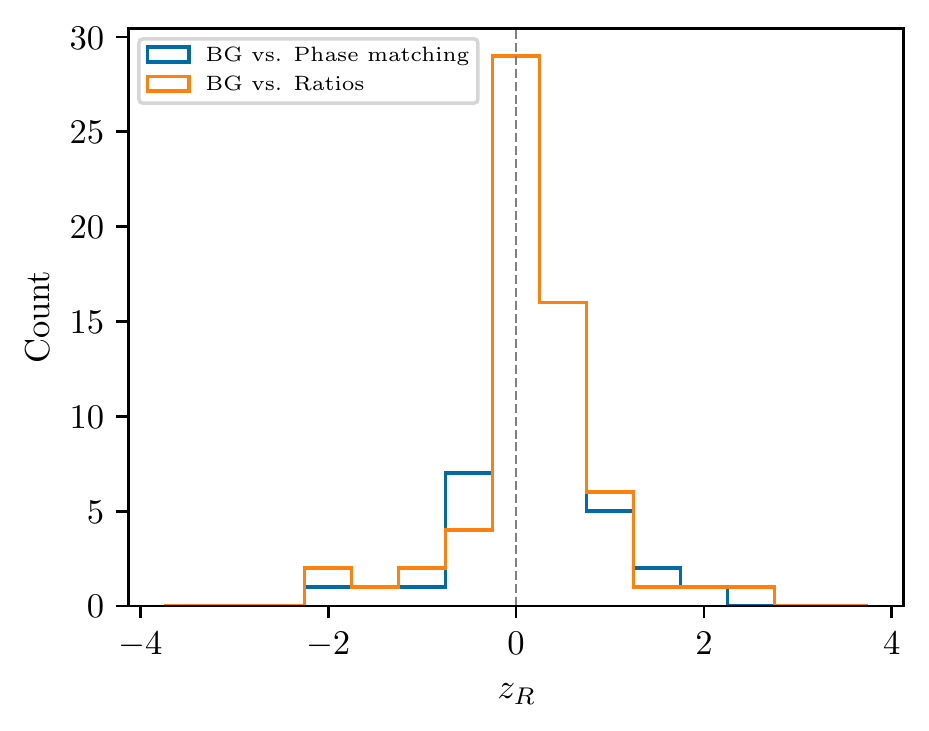}
    \includegraphics[width=.45\textwidth, trim=.25cm .25cm .25cm .15cm,clip]{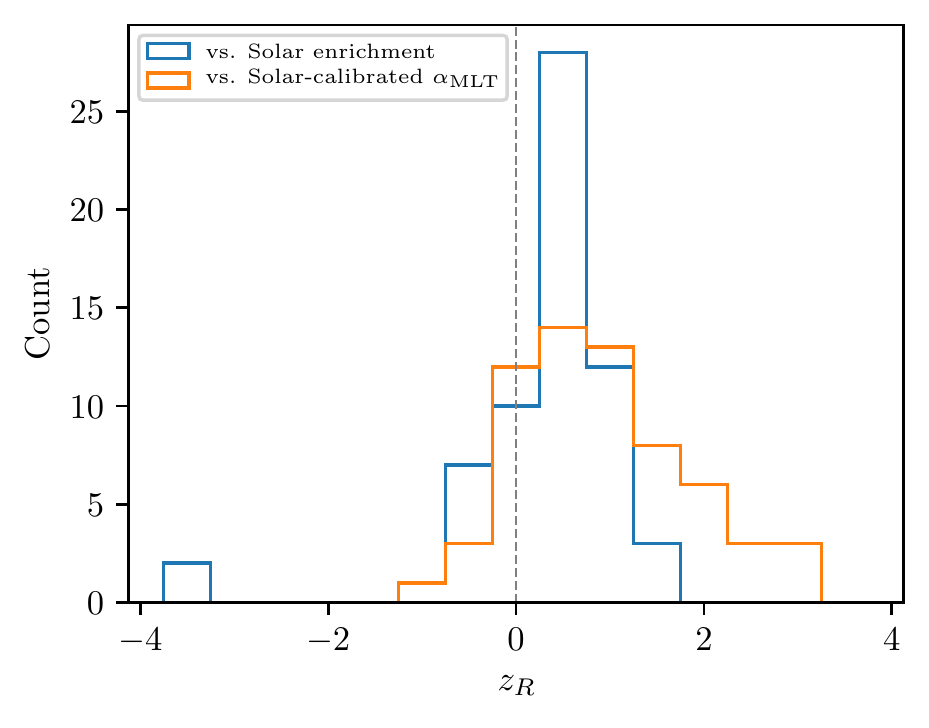}
    \includegraphics[width=.45\textwidth, trim=.25cm .25cm .25cm .15cm,clip]{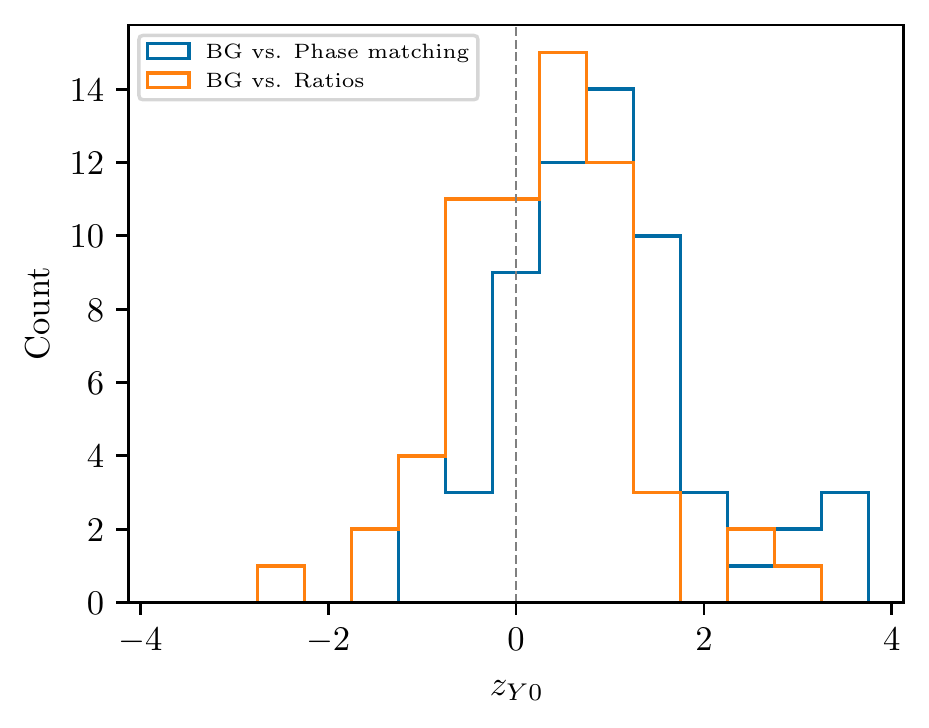}
    \caption{Distributions of $z$-scores used for different hypothesis tests. \textbf{Top}: Normalised differences in posterior radii between different surface corrections under a solar-like linear enrichment law. \textbf{Middle}: Normalised differences in posterior radii from different conditional distributions (compared to the full grid), with a \bg\ surface correction.\textbf{Bottom}: Normalised differences in initial helium abundances for different surface corrections, compared to that of \bg.}
    \label{fig:difft}
\end{figure}

We see that most of these \(t\)-tests do not return significant results
by our predetermined criterion, indicating that the relative effects of
the three surface terms that we have examined do not matter, on average.
Except for the initial helium abundance, even the most significant set
of normalised differences (between radii inferred under a solar-like
enrichment law) has a distribution whose mean differs from zero only
barely discernibly (top panel of \cref{fig:difft}). This is consistent
with the findings of \citet{compton_surface_2018} and
\citet{basu_robustness_2018}. We have generalised the former by
including nonparametric surface corrections in the analysis, and the
latter by demonstrating this on a larger, statistical sample.

By contrast, these constraints on model selection are known to
significantly affect the estimated stellar properties. To compare the
size of these effects, we perform a similar set of tests for systematic
differences between inferred parameters returned from each of the above
choices of modelling constraints vs.~from the full grid, keeping the
choice of surface correction consistent. We show the results of these
tests in \cref{tab:t2}. Again, each entry in the table consists of two
numbers, which are \(p\)-values for comparisons between a linear
enrichment law vs.~the full grid, and between a solar-calibrated mixing
length vs.~the full grid. We see that these differences in model
selection contraints yield normalised differences that are clearly
nonzero on average (which we illustrate in the middle panel of
\cref{fig:difft}). The exception to this is the stellar age, which
apparently cannot discriminate between these modelling choices. We
conclude that the properties of the underlying stellar models have a
more significant impact on most inferred global parameters than does the
choice of surface correction used when matching them to observations, in
agreement with prior work.

\renewcommand\subrow{{\footnotesize$\begin{array}{c}\text{full vs. 1} \\ \text{full vs. 2}\end{array}$}}

\begin{longtable}[]{@{}lrccc@{}}
\caption{\(t\)-test \(p\)-values for normalised differences between
model selection restrictions, numbered as in \cref{tab:t1}.
\label{tab:t2}}\tabularnewline
\toprule
\begin{minipage}[b]{0.10\columnwidth}\raggedright
Quantity\strut
\end{minipage} & \begin{minipage}[b]{0.13\columnwidth}\raggedleft
\strut
\end{minipage} & \begin{minipage}[b]{0.21\columnwidth}\centering
\bg\strut
\end{minipage} & \begin{minipage}[b]{0.21\columnwidth}\centering
Phase\strut
\end{minipage} & \begin{minipage}[b]{0.21\columnwidth}\centering
Ratios\strut
\end{minipage}\tabularnewline
\midrule
\endfirsthead
\toprule
\begin{minipage}[b]{0.10\columnwidth}\raggedright
Quantity\strut
\end{minipage} & \begin{minipage}[b]{0.13\columnwidth}\raggedleft
\strut
\end{minipage} & \begin{minipage}[b]{0.21\columnwidth}\centering
\bg\strut
\end{minipage} & \begin{minipage}[b]{0.21\columnwidth}\centering
Phase\strut
\end{minipage} & \begin{minipage}[b]{0.21\columnwidth}\centering
Ratios\strut
\end{minipage}\tabularnewline
\midrule
\endhead
\begin{minipage}[t]{0.10\columnwidth}\raggedright
Mass\strut
\end{minipage} & \begin{minipage}[t]{0.13\columnwidth}\raggedleft
\subrow\strut
\end{minipage} & \begin{minipage}[t]{0.21\columnwidth}\centering
\(\begin{array}{c}0.0021 \\ 3.40 \times 10^{-10}\end{array}\)\strut
\end{minipage} & \begin{minipage}[t]{0.21\columnwidth}\centering
\(\begin{array}{c}5.13 \times 10^{-10} \\ 9.26 \times 10^{-9}\end{array}\)\strut
\end{minipage} & \begin{minipage}[t]{0.21\columnwidth}\centering
\(\begin{array}{c}2.15 \times 10^{-6} \\ 1.07 \times 10^{-9}\end{array}\)\strut
\end{minipage}\tabularnewline
\begin{minipage}[t]{0.10\columnwidth}\raggedright
Radius\strut
\end{minipage} & \begin{minipage}[t]{0.13\columnwidth}\raggedleft
\subrow\strut
\end{minipage} & \begin{minipage}[t]{0.21\columnwidth}\centering
\(\begin{array}{c}0.0030 \\ 1.60 \times 10^{-11}\end{array}\)\strut
\end{minipage} & \begin{minipage}[t]{0.21\columnwidth}\centering
\(\begin{array}{c}1.21 \times 10^{-9} \\ 2.61 \times 10^{-10}\end{array}\)\strut
\end{minipage} & \begin{minipage}[t]{0.21\columnwidth}\centering
\(\begin{array}{c}3.38 \times 10^{-6} \\ 1.05 \times 10^{-11}\end{array}\)\strut
\end{minipage}\tabularnewline
\begin{minipage}[t]{0.10\columnwidth}\raggedright
Age\strut
\end{minipage} & \begin{minipage}[t]{0.13\columnwidth}\raggedleft
\subrow\strut
\end{minipage} & \begin{minipage}[t]{0.21\columnwidth}\centering
\(\begin{array}{c}0.9088 \\ 0.1540\end{array}\)\strut
\end{minipage} & \begin{minipage}[t]{0.21\columnwidth}\centering
\(\begin{array}{c}0.6442 \\ 0.3197\end{array}\)\strut
\end{minipage} & \begin{minipage}[t]{0.21\columnwidth}\centering
\(\begin{array}{c}0.6572 \\ 0.0015\end{array}\)\strut
\end{minipage}\tabularnewline
\begin{minipage}[t]{0.10\columnwidth}\raggedright
\(Y_0\)\strut
\end{minipage} & \begin{minipage}[t]{0.13\columnwidth}\raggedleft
\subrow\strut
\end{minipage} & \begin{minipage}[t]{0.21\columnwidth}\centering
\(\begin{array}{c}1.15 \times 10^{-5} \\ 8.36 \times 10^{-7}\end{array}\)\strut
\end{minipage} & \begin{minipage}[t]{0.21\columnwidth}\centering
\(\begin{array}{c}1.56 \times 10^{-8} \\ 1.64 \times 10^{-7}\end{array}\)\strut
\end{minipage} & \begin{minipage}[t]{0.21\columnwidth}\centering
\(\begin{array}{c}2.26 \times 10^{-7} \\ 6.04 \times 10^{-6}\end{array}\)\strut
\end{minipage}\tabularnewline
\bottomrule
\end{longtable}

An exception to this is the initial helium abundance, where we find that
the method of phase matching yields significantly different values
compared to the other two surface corrections, even when the full grid
is employed (bottom panel of \cref{fig:difft}). Comparing the
\(p\)-values in \cref{tab:t1,tab:t2}, we find that the size of this
effect is comparable to that induced by different model selection
constraints. Therefore, we conclude that inference of the helium
abundance in particular, unlike other model parameters, is sensitive to
the choice of surface correction, and therefore methodology-dependent in
this respect.

\hypertarget{differences-in-correlations-between-model-parameters}{%
\subsection{Differences in correlations between model
parameters}\label{differences-in-correlations-between-model-parameters}}

As can be seen in \cref{fig:corner}, the posterior distributions for
various model parameters are correlated with each other \citep[see
e.g.][]{lebreton_alacarte_2014}. We are in particular concerned with the
dependences of fundamental stellar properties on the choice of initial
helium abundance \(Y_0\) and the mixing-length parameter \amlt, since
these model parameters directly affect the other inferred properties,
and are also not otherwise well-constrained by direct observations. As
the stellar mass is the primary quantity of astrophysical interest in
most cases for main-sequence stars, we focus on its correlation with
these parameters in particular.

In principle, we could characterise this in terms of the joint second
cumulants (i.e.~covariances) of the posterior distribution. However,
these do not lend themselves easily to physical interpretation. An
alternative would be to find the principal axes of the joint posterior
distribution (i.e.~eigenvectors of the covariance matrix), but this
requires that the posterior distributions be symmetric and not curved,
which we find to not always be the case.

\begin{figure}
\centering
\includegraphics{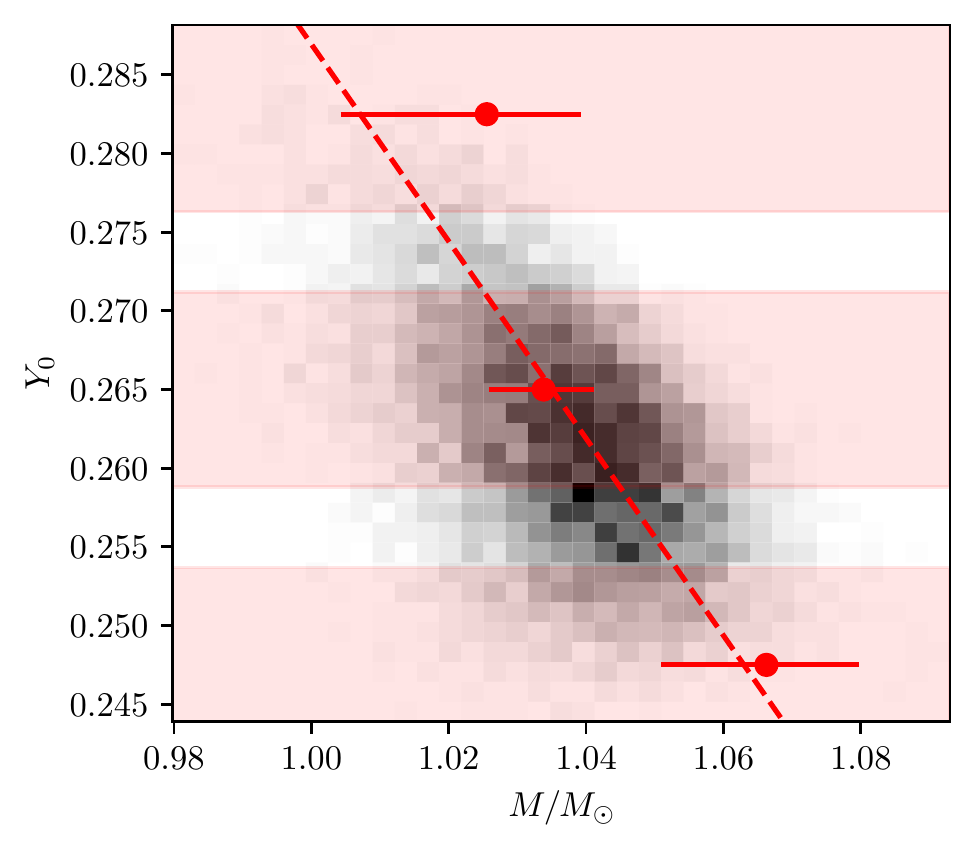}
\caption{Construction of the mass-helium correlation parameter,
\(\mathrm d M / \mathrm d Y\) (\cref{eq:gradient}), illustrated with 16
Cyg B. We show in the background (in greyscale) the marginal joint
distribution between \(M\) and \(Y_0\) (same as the top panel of
\cref{fig:corner}). The pink shaded regions show the 1\(\sigma\) regions
of the Gaussian weights, \cref{eq:conditional2}, associated with each
corresponding point, which shows the conditional median and
\(\pm1\sigma\) confidence intervals in the inferred masses associated
with that value of \(Y_0\). Together, these points imply that the
posterior masses are correlated with the initial helium abundance, which
we characterise with a gradient parameter,
\(\mathrm d M / \mathrm d Y_0\), that is shown with the dashed
line.\label{fig:gradient}}
\end{figure}

We opt instead to quantify these correlations via the weighted averages
of the gradient of the conditional posterior mean of one quantity with
respect to another. For example, the conditional posterior mass
\(M|_{Y_0 = Y}\) is in general a function of \(Y\), and we would like to
take the average of the derivative of this function over the values of
\(Y\) permitted by the grid. In principle, we should do this by
approximating the formal expression \begin{equation}
    {\mathrm d M \over \mathrm d Y_0} \sim \int\mathrm dY f(Y) {\partial \over \partial Y} \int M(x) p(x | Y_0 = Y)\ \mathrm d^n x,\label{eq:gradient}
\end{equation} where \(f\) is the uniform distribution over our input
range of, say, the initial helium abundances. Rather than evaluating
these derivatives directly, we instead compute \(M|_{Y_0 = Y_{0,i}}\)
using the discretised approximations \cref{eq:mean,eq:conditional} for
several values of \(Y_{0, i}\), and fit a linear model weighted by the
posterior uncertainties. Our final correlation parameter is the gradient
of this line. We illustrate this procedure in \cref{fig:gradient}. We
apply the same procedure to find the average derivative of the posterior
mass with respect to \amlt.

\begin{figure}[htbp]
    \centering
    \includegraphics[width=.45\textwidth, trim=.25cm .75cm .25cm .15cm,clip]{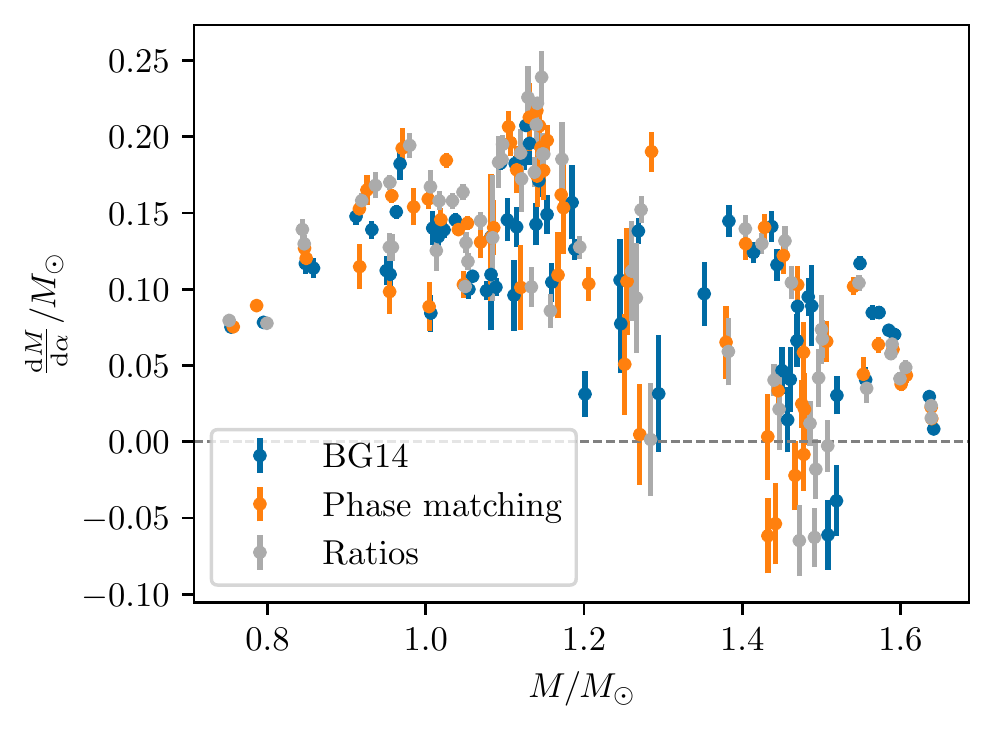}
    \includegraphics[width=.45\textwidth, trim=.25cm .25cm .25cm .15cm,clip]{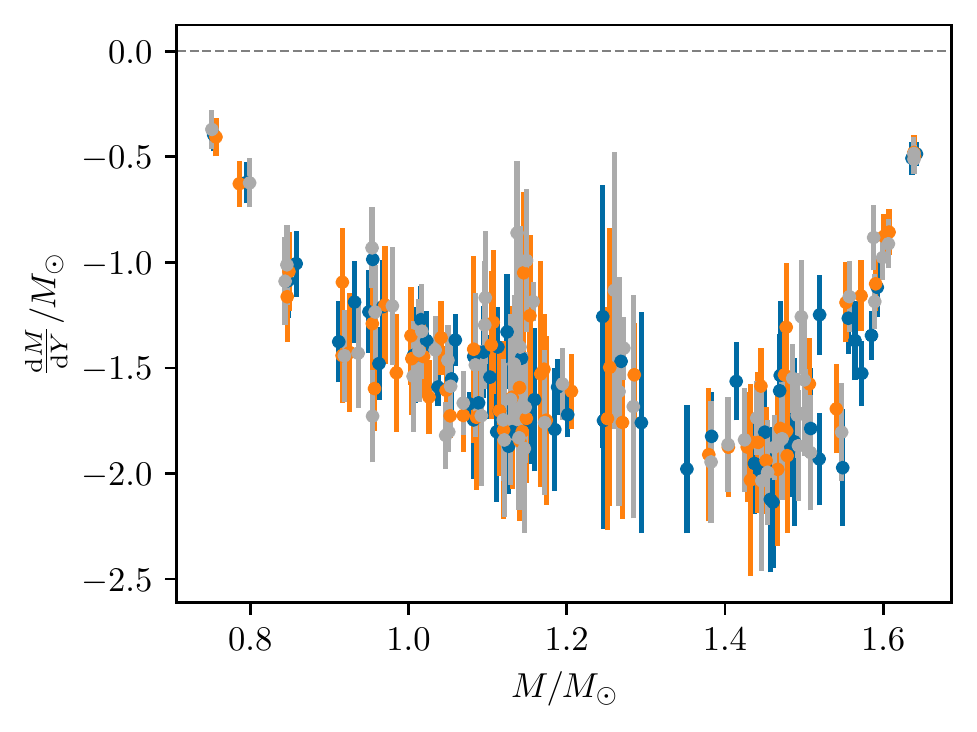}
    \caption{Correlations between posterior mass and the mixing length parameter (top) and initial helium abundance (bottom), for different choices of surface correction.}
    \label{fig:ms}
\end{figure}

We show these gradients in \cref{fig:ms}, as computed with respect to
each of the surface corrections we have considered. Broadly speaking, it
is difficult to discern any visible structure in the differences between
surface corrections, which again suggests that these correlations are a
property of the stellar models in the grid rather than on the surface
terms used to match them to observational constraints. However, we would
like to make this claim more rigorous.

We therefore perform a similar procedure to that in the previous
section: for a given pair of surface terms, we construct pairwise
differences of the mass-helium correlation parameters for each star, and
test the null hypothesis that they are distributed around zero. A
significant deviation from this null hypothesis indicate that the two
surface corrections result in different systematic correlations. We
repeat this procedure for both of the constrained model selection
scenarios (solar mixing length and solar enrichment law) as in the
previous section.

\begin{figure*}[htbp]
    \centering
    \includegraphics[width=.45\textwidth, trim=.15cm .75cm .25cm .15cm,clip]{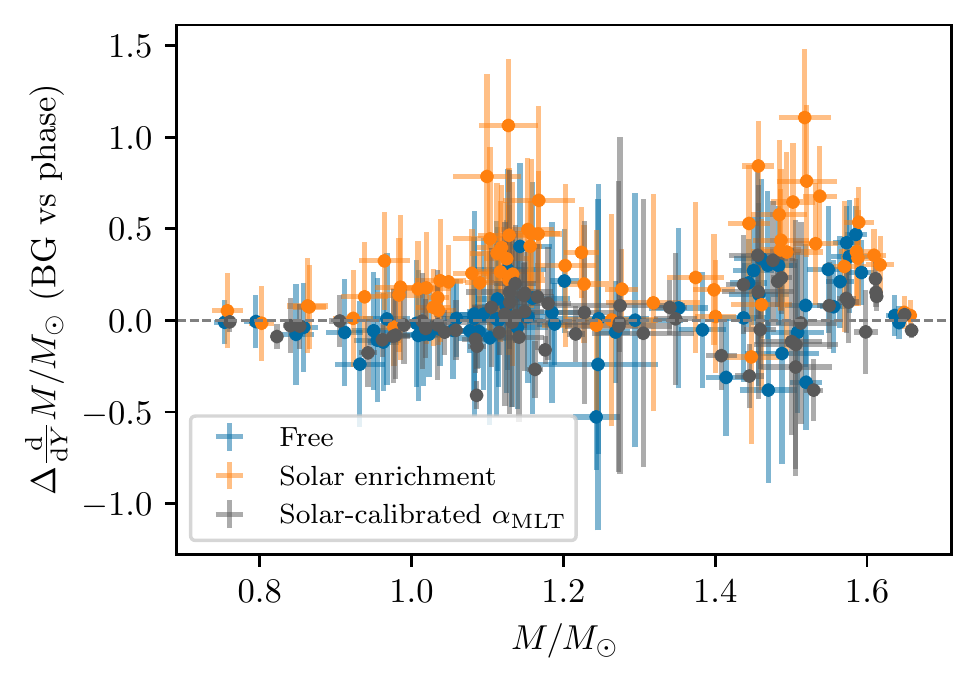}
    \includegraphics[width=.45\textwidth, trim=.15cm .75cm .25cm .15cm,clip]{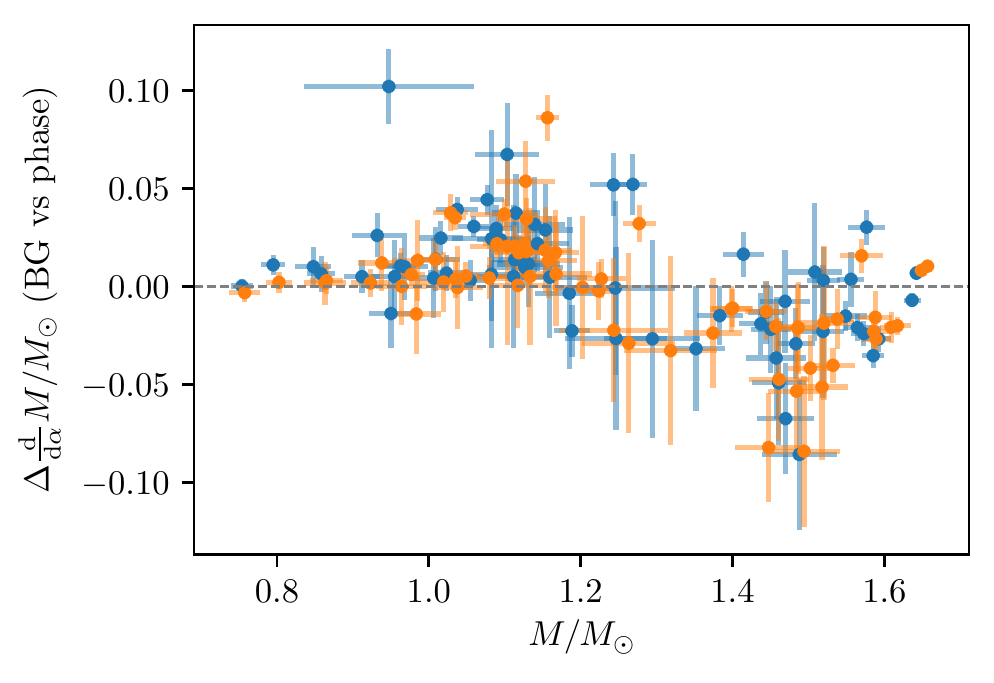}
    \includegraphics[width=.45\textwidth, trim=.15cm .75cm .25cm .15cm,clip]{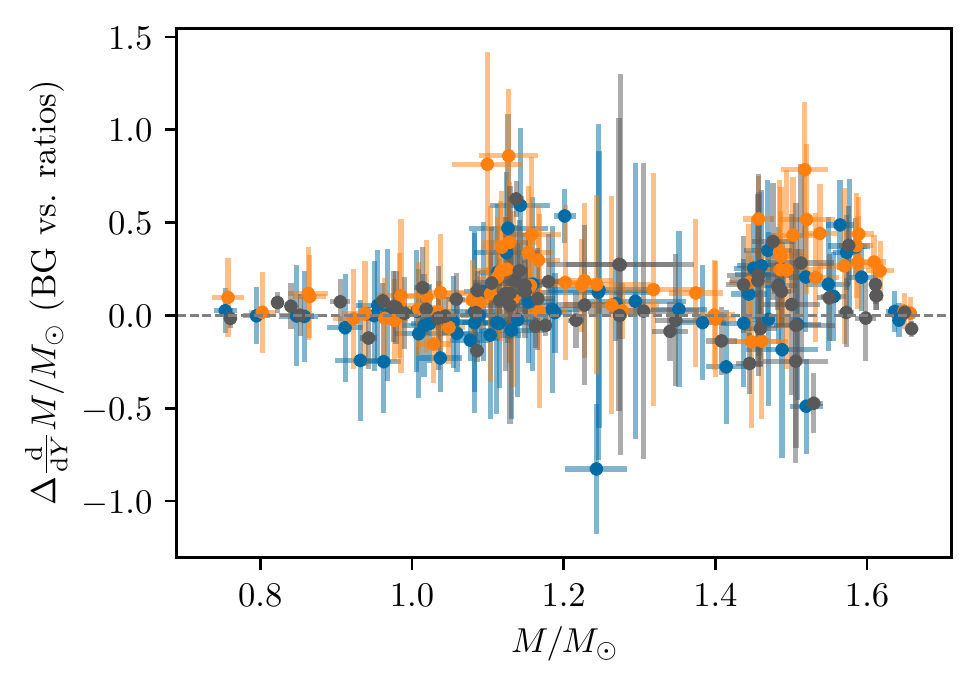}
    \includegraphics[width=.45\textwidth, trim=.15cm .75cm .25cm .15cm,clip]{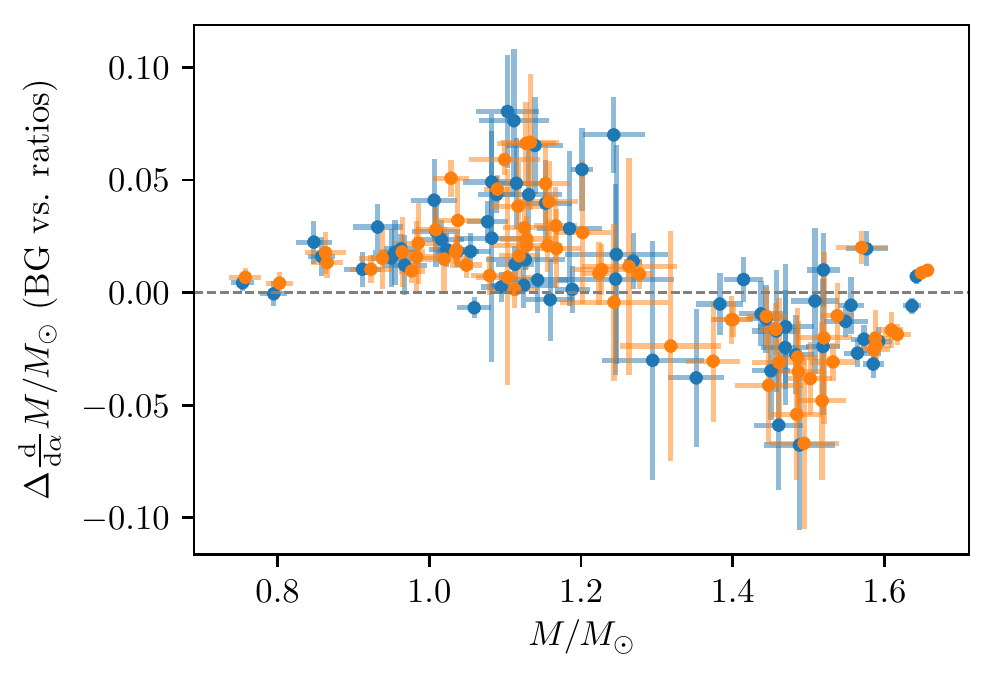}
    \includegraphics[width=.45\textwidth, trim=.15cm .25cm .25cm .15cm,clip]{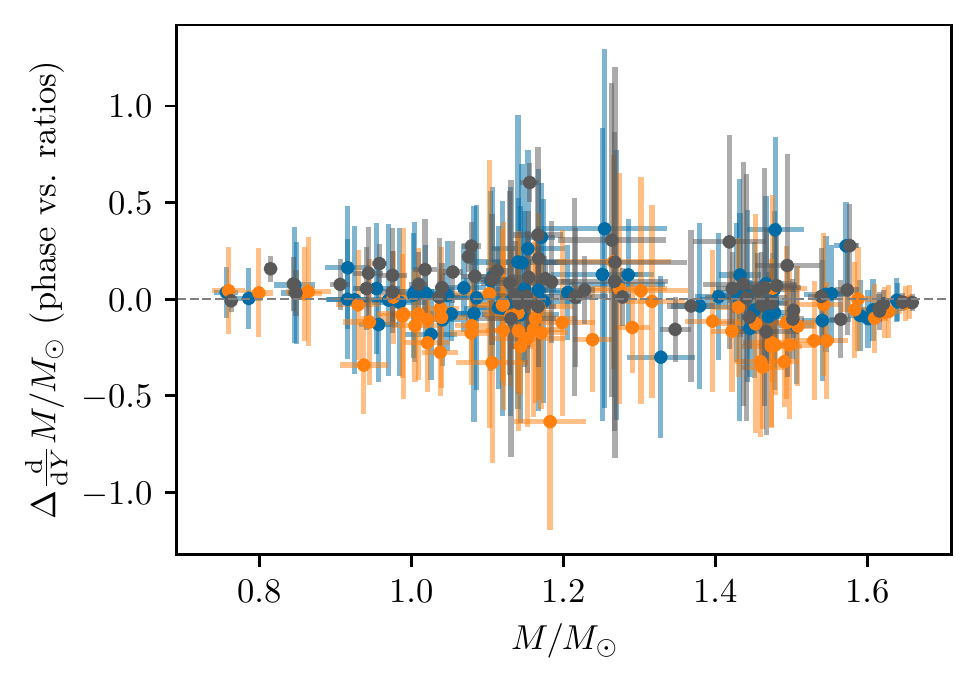}
    \includegraphics[width=.45\textwidth, trim=.15cm .25cm .25cm .15cm,clip]{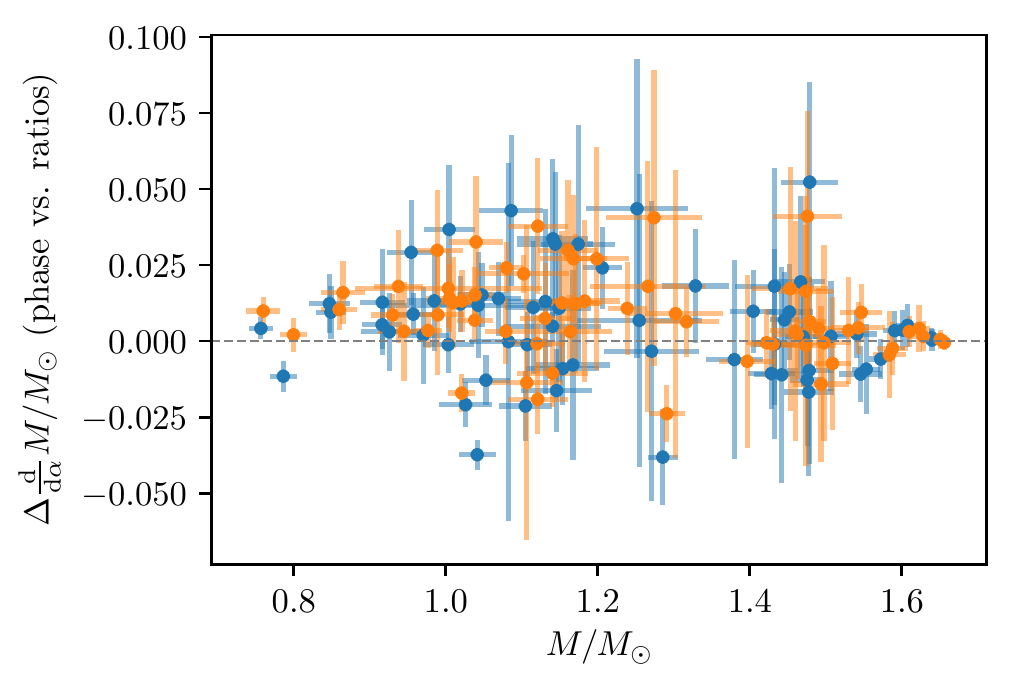}
    \caption{Differences in the mass-helium (left column) and mass-mixing (right column) correlation parameters between various pairs of surface corrections. Colours represent different model selection constraints. From top to bottom, we compare the \bg\ correction to the method of phase matching (in the sense of $\epsilon - \text{BG}$); \bg\ correction to the method of separation ratios; and phase matching to separation ratios.}
    \label{fig:diffnuisance}
\end{figure*}

\begin{longtable}[]{@{}lccc@{}}
\caption{\(t\)-test \(p\)-values for differences in
\(\mathrm d M /\mathrm d Y\)}\tabularnewline
\toprule
Surface corrections & Full Grid & Solar Enrichment & Solar
\(\amlt\)\tabularnewline
\midrule
\endfirsthead
\toprule
Surface corrections & Full Grid & Solar Enrichment & Solar
\(\amlt\)\tabularnewline
\midrule
\endhead
\bg~vs.~phase & \(0.1638\) & \(9.04 \times 10^{-9}\) &
\(0.6399\)\tabularnewline
\bg~vs.~ratios & \(0.4576\) & \(8.17 \times 10^{-7}\) &
\(0.0943\)\tabularnewline
phase vs ratios & \(0.8422\) & \(8.61 \times 10^{-5}\) &
\(0.0084\)\tabularnewline
\bottomrule
\end{longtable}

We show these pairwise differences in \cref{fig:diffnuisance}, colouring
points by different model selection constraints. Even visually, it is
clear that the imposition of a solar-like linear enrichment law affects
each of the surface corrections differently. That is to say, while the
correlation between the posterior masses and initial helium abundances
depends primarily on the properties of the stellar models, the precise
nature of this has a secondary dependence on the choice of surface term.
We should qualify that this is in any case a second-order effect: it
pertains to the inferred covariances, and not to the global properties
per se.

A similar analysis for mass-\amlt~correlation parameters does not reveal
any significant dependence on the surface correction (with a minimum
\(p\)-value of 0.007 among all of the combinations that we examined).
However, \cref{fig:diffnuisance} suggests that the correlations with
respect to \(\amlt\) returned by the \bg~correction differ from those
returned by our other two surface corrections in a mass-dependent
manner. Nonetheless, this is once again only a second-order effect.
While interesting from a purely statistical perspective, these are
likely to be of little consequence for most practical applications.
However, in conjunction with our discussion in the previous section,
this serves to illustrate again that estimation of the helium abundance
in particular, from asteroseismic constraints on evolutionary models,
appears to be a highly methodology-dependent affair. While this already
known to be the case for other aspects of stellar modelling, we have at
least demonstrated that such methodological considerations must also
extend to the choice of treatment for the surface term.

\hypertarget{core-diagnostics-from-separation-ratios}{%
\subsection{Core diagnostics from separation
ratios}\label{core-diagnostics-from-separation-ratios}}

\begin{figure*}[htbp]
    \centering
    \annotate{\includegraphics[width=.475\textwidth, trim=0cm .25cm 0cm .15cm,clip]{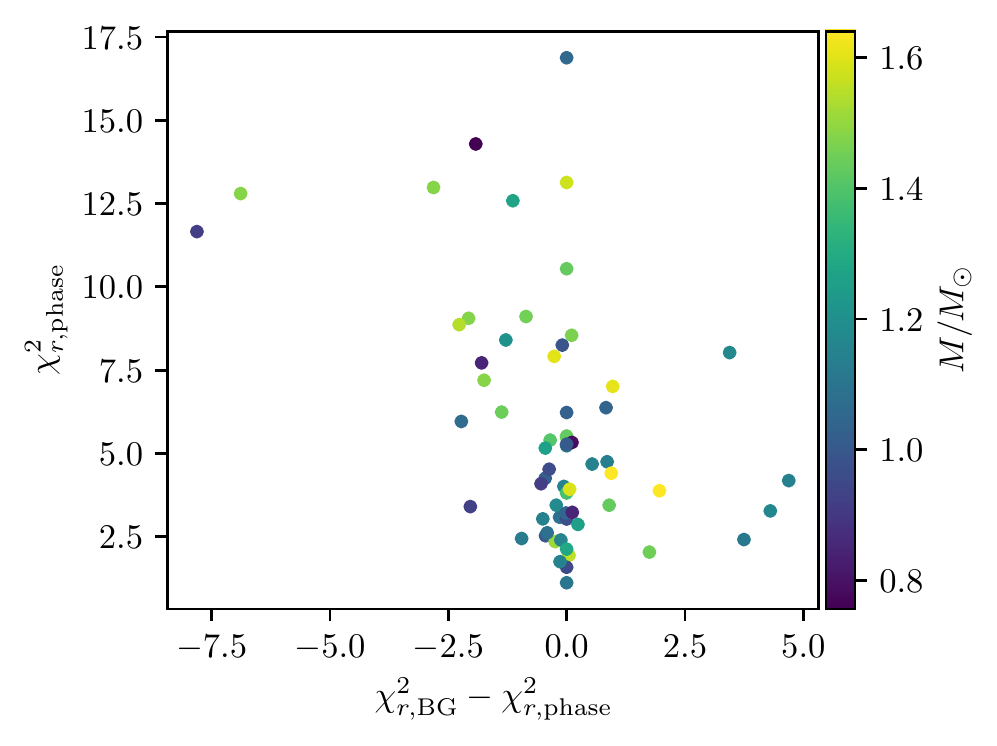}}{\node at (.22, .22){\textbf{(a)}};}
    \annotate{\includegraphics[width=.455\textwidth, trim=0cm .25cm 0cm .15cm,clip]{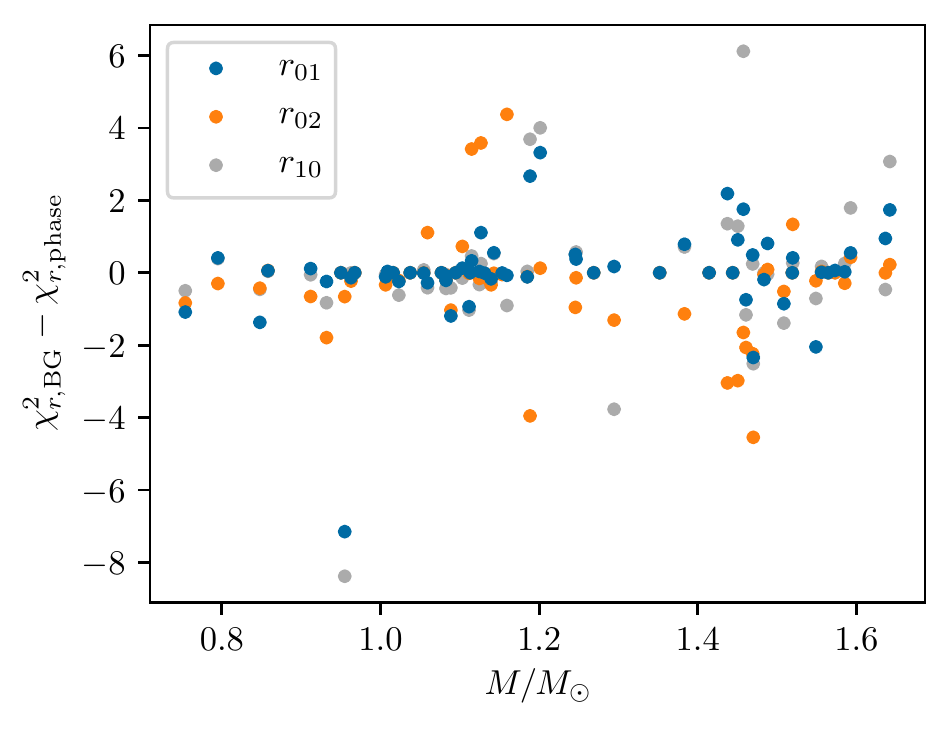}}{\node at (.22, .22){\textbf{(b)}};}
    \annotate{\includegraphics[width=.45\textwidth, trim=.25cm .25cm .25cm .0cm,clip]{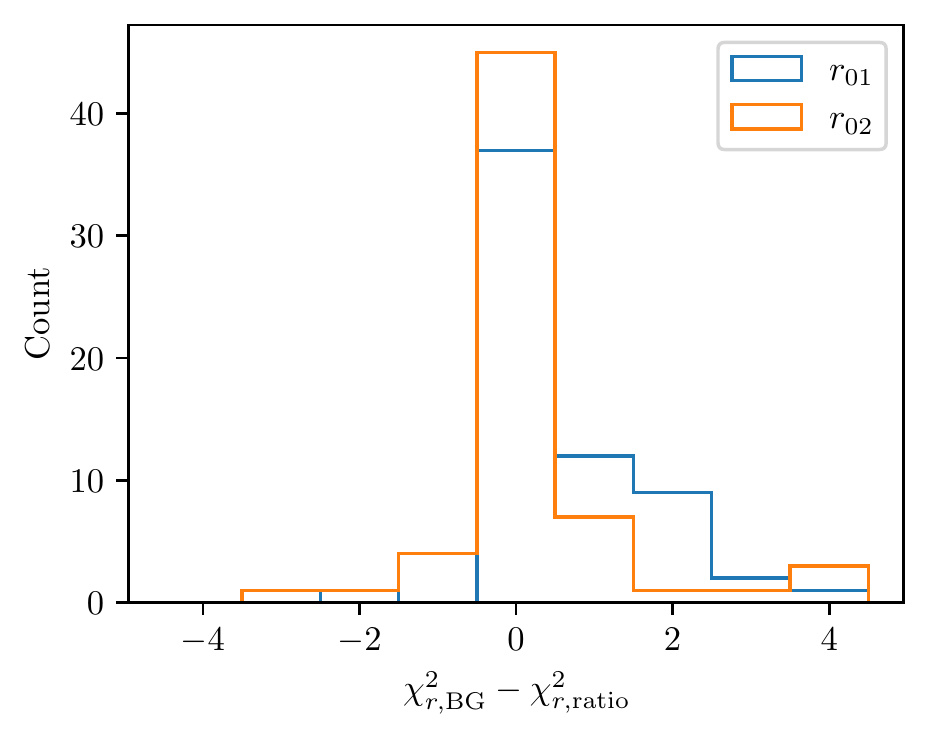}}{\node at (.17, .22){\textbf{(c)}};}
    \annotate{\includegraphics[width=.435\textwidth, trim=.55cm .25cm .25cm .0cm,clip]{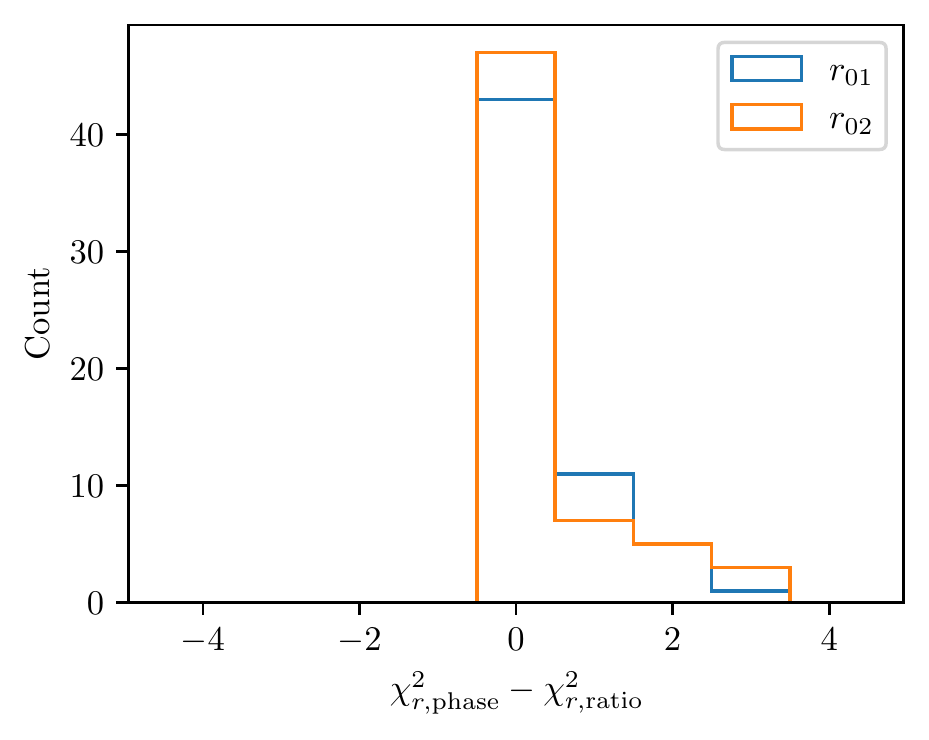}}{\node at (.17, .22){\textbf{(d)}};}
    \caption{Core diagnostics from separation ratios. \textbf{(a)} Differences between $\chi^2_\text{ratio}$ for best-fitting models returned from the \bg\ and phase-matching methods for each star, shown with $\chi^2_\text{ratio}$ for the phase-matching method on the vertical axis.\textbf{(b)} These differences broken up by $r_{01}$, $r_{10}$, and $r_{02}$; the best-fitting model masses of the phase-matching method are shown on the horizontal axis. \textbf{(c)} Distribution of $\chi^2_{r_{01}}$ and $\chi^2_{r_{02}}$ of the best-fitting \bg\ models compared to the best-fitting separation-ratio models. \textbf{(d)} Distribution of $\chi^2_{r_{01}}$ and $\chi^2_{r_{02}}$ of the best-fitting phase-matching models compared to the best-fitting separation-ratio models.}
    \label{fig:diffratios}
\end{figure*}

One generic problem with all three treatments of the surface term we
have considered is that while poor fits (high costs) imply modelling
error in the interior structure, the converse may not be true; good fits
(low costs) do yield precise estimates of global stellar properties, but
they may not necessarily also yield models that match the interior
structure of the stars. For instance, structural inversion of the sound
speed profiles of 16 Cyg A and B \citep{bellinger_model_2017} indicates
that stellar models matched to observed frequencies via the \bg~surface
correction are known to possess significant internal differences from
the structures of the actual stars. On the other hand,
\citet{bellinger_model_2017} also demonstrate that in these cases, the
frequency ratios serve as diagnostics of the interior structure, in the
sense that for good-fitting models with respect to the \bg~correction,
discrepancies with the observed frequency separation ratios still
indicate discrepancies in the interior structure.

Given this to be the case, we assume that for otherwise good-fitting
models with respect to the \bg~and phase-matching surface corrections,
the separation ratios may serve as diagnostics of structural mismatches
in the interior. For each of these two methods, let
\(\chi^2_{r, \text{corr}}\) be the value of \(\chi^2_\text{ratio}\)
(\autoref{sec:ratios}) associated with the best-fitting model returned
by that surface correction. The difference between the two then serves
as an informal diagnostic of which method performs better at recovering
the internal structure of the actual star; it may be loosely interpreted
as a likelihood ratio with respect to being constrained by the
separation ratios.

We show these differences on the horizontal axis of
\cref{fig:diffratios}a, with the separation-ratio cost function of the
best-fitting phase-matching model on the vertical axis. Broadly
speaking, most of these values are situated very close to zero,
indicating that, as far as the separation-ratio diagnostic is concerned,
the best-fitting models selected by both methods are equally good (or
equally bad) matches to the interior of the actual star. In
\cref{fig:diffratios}(b), we break this down by each of these separation
ratios considered separately. This reveals no obvious dependence on
degree or on the stellar mass. Since dipole modes probe deeper into the
stellar interior than do quadrupole modes, this appears to indicate that
any structural mismatches between the best-fitting models selected by
the two methods are not preferentially localised.

We also compare the best-fitting model for each method with that
selected by using the separation ratios directly; we show the results
for the \bg~correction in \cref{fig:diffratios}c, and those for the
phase-matching method in \cref{fig:diffratios}d.~We see that the
phase-matching method returns best-fitting models that uniformly diverge
more from the implied interior structure than those selected directly by
the separation-ratio constraint. Surprisingly, however, the
\bg~correction appears to sacrifice accuracy in matching the dipole mode
separation ratios, in exchange for matching the quadrupole mode ratios
equally well, on average, as the best-fitting models obtained from
fitting the separation ratios directly.

\begin{figure}
\centering
\includegraphics{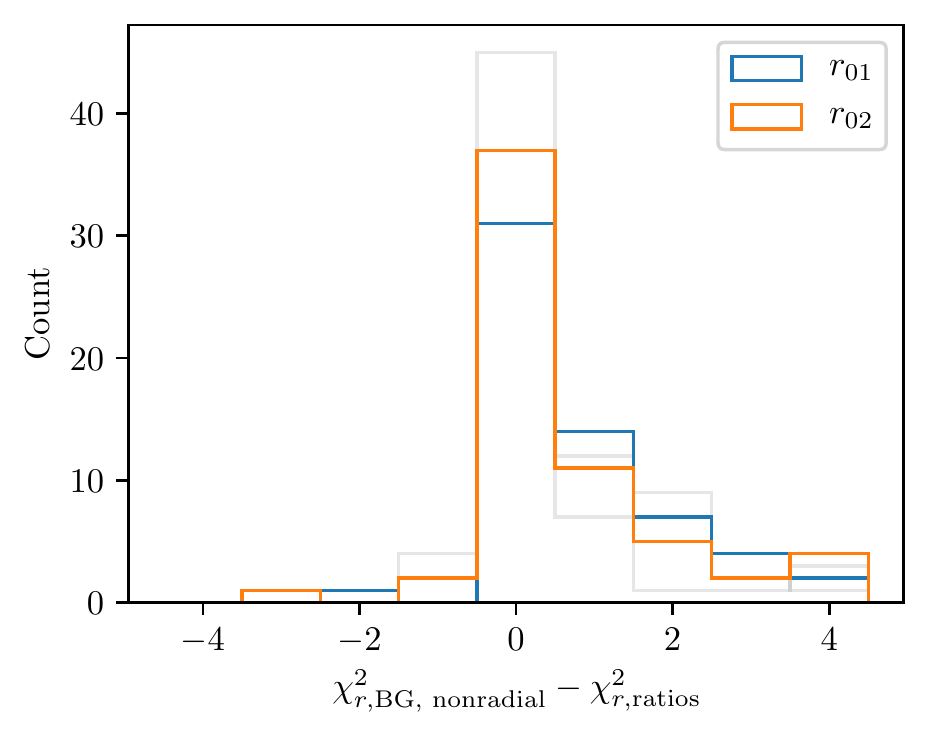}
\caption{Same as \cref{fig:diffratios}c (shown with faint lines in the
background), except with the inclusion of nonradial modes in the
procedure for fitting the parameters of the \bg~surface correction. On
average, this yields \(r_{01}\) and \(r_{02}\) that both deviate more
from those implied by the observed modes.\label{fig:nonradial}}
\end{figure}

This degree dependence could possibly be a property of our
implementation of the \bg~correction, where we have used only radial
modes in fitting the free parameters (see \autoref{sec:bgcost}). To
determine if this is the case, we repeated our grid search using radial,
dipole, and quadrupole modes to fit the \bg~parameters \(a_{-1}\) and
\(a_3\), rather than just the radial modes. We show the same
log-likelihood-ratios as \cref{fig:diffratios}c in \cref{fig:nonradial}.
While including the nonradial modes does appear to alleviate the
preference for matching \(r_{02}\) better than \(r_{01}\), it does so at
the expense of overall performance. On average, this procedure appears
to select models with internal structures that deviate more from the
actual stars than those selected by using only radial modes in fitting
the \bg~coefficients, or, for that matter, by using the phase-matching
method. We do not have a good explanation for this.

The coefficients \(a_{-1}, a_{3}\) of the \bg~parametrisation are found
independently for each model by minimising the cost function of
\cref{eq:bgcost}. When modelling evolved stars, it often the case that
only radial modes are included in this cost function when constraining
the \bg~coefficients, since the action of the surface term on mixed
modes is nontrivial \citep[see][ and
\autoref{sec:coupling}]{ball_surface_2018}, whereas on the main
sequence, where all observed modes can be safely assumed to be
p-dominated, it is more typical to use modes of all observed degrees to
constrain these coefficients. This result suggests that,
counterintuitively, the use of only radial modes in this procedure
returns better structural matches (as diagnosed by surface-independent
separation ratios) even for main-sequence stars, where mode mixing is
not ordinarily a concern.

\hypertarget{red-giants}{%
\section{Red giants}\label{red-giants}}

Unlike the case for main-sequence stars, extant comparisons of surface
corrections for red giants have so far investigated only parametric
corrections, owing to concerns about the validity of nonparametric
methods in the presence of mode mixing
\citep[e.g.][]{ball_surface_2017}. We would nonetheless like to
investigate how they hold up to parametric methods in any case. We
describe rudimentary modifications to mitigate the effects of mode
mixing in \autoref{sec:coupling}. Otherwise, we used the same modelling
procedure to infer the parameters and posterior distributions of 28
first-ascent red-giant branch (RGB) stars in the open cluster NGC~6791.
We use the same spectroscopic properties, measured mode frequencies,
stellar model grids, and model frequency calculations as
\citet{mckeever_helium_2019}.

\hypertarget{model-grid-1}{%
\subsection{Model grid}\label{model-grid-1}}

We utilized the pre-existing grid of models from
\citet{mckeever_helium_2019} to study these relations for red giants.
This grid of models was specifically designed to model the open cluster
NGC~6791, and is thus more limited in mass range than our main sequence
models. The grid was constructed with \mesa~r10108, using an Eddington
gray atmosphere, the mixing-length prescription of
\citet{cox_principles_1968}, the gravitational settling prescription of
\citet{thoul_element_1994}, the elemental abundances of
\citet{grevesse_standard_1998}, and with a small amount of convective
overshoot (\(f_\text{ov}=0.016\)). The grid was discretely sampled at
\(M/M_\Sun \in [1.1, 1.25]\) in steps of \(0.01 M_\Sun\),
\(Y_0 \in [0.28, 0.32]\) in steps of \(0.01\) (with two additional
intermediate grid points bracketing the central value),
\([\text{Fe/H}]_0 \in [0.25, 0.45]\) in steps of \(0.05\), and
\(\amlt \in [1.7, 2.1]\) in steps of \(0.1\). This discrete sampling
meant that we were unfortunately unable to restrict our model selection
to e.g.~a fixed \(\mathrm d Y/\mathrm d Z\) relation, as we did for our
main-sequence stars with Sobol sampling. More details about the grid can
be found in the original paper.

\hypertarget{effects-of-mixed-mode-coupling}{%
\subsection{Effects of mixed-mode
coupling}\label{effects-of-mixed-mode-coupling}}

\label{sec:coupling}

The validity of both of the nonparametric methods that we have examined
relies on a partial-wave eigenvalue equation depending only on the
p-mode radial order \(n_p\). They therefore apply only to uncoupled
p-modes, i.e.~in the absence of g-mode coupling. In evolved stars,
however, this coupling perturbs the frequencies of nonradial p-modes
that are close to resonance with internal g-modes of the same degree
through the avoided-crossing phenomenon.

The most p-dominated mixed modes returned from observed oscillation
power spectra are known to correspond to model mixed modes that achieve
maximum visibility, as parameterised by the inverse inertia
\citep[e.g.][]{dupret_amplitudes_2009, aertsbook}. As demonstrated in
\citet{ong_semianalytic_2020}, these occur close to the isolated
\(\pi\)-mode frequencies \citep[in the sense
of][]{aizenman_avoided_1977}, which are constructed to yield pure
pressure-mode oscillations with no mixing from interior gravity waves.
The isolated eigenvalue problem for \(\pi\)-modes admits a similar
characterisation to uncoupled p-modes in terms of inner and outer
partial waves, despite being technically inequivalent. In principle,
were these \(\pi\) modes observationally accessible, they could be used
to construct separations ratios in the same manner, constrained against
those constructed from model \(\pi\)-modes. However, the measured mode
frequencies (which cannot be isolated) will necessarily differ from
those of the true \(\pi\)-modes owing to mode bumping from the
avoided-crossing phenomenon
\citep[e.g.][]{deheuvels_constraints_2011, mosser_probing_2012}. These
deviations are most pronounced with the dipole modes, where the coupling
is strongest and the g-mode period spacing is largest
\citep[e.g.][]{pincon_probing_2020, jiang_variations_2020}; these dipole
modes are in turn used in the computatation of all of the standard
separation ratios.

As such, we do not use the dipole separation ratios \(r_{01}\) and
\(r_{10}\) as a constraint in modelling the red giant sample. On the
other hand, frequency separations \(\delta\nu_{02}\) and separation
ratios \(r_{02}\) are known to have been successfully derived from red
giants, and employed to obtain good agreement with stellar modelling,
where these quantities are computed from the most p-dominated quadrupole
mixed modes
\citep[cf.][]{huber_asteroseismology_2010, white_calculating_2011, broomhall_prospects_2014}.
This is possible because the \(\pi\)-\(\gamma\) coupling for quadrupole
modes is much weaker than for dipole modes; for further discussion of
this, see \autoref{ap:coupling}. Consequently, we do assume that the
observed quadrupole modes are representative of the \(\pi\) mode
frequencies. Assuming this to be the case, we computed \(r_{02}\) using
the modified definition \citep[see][Eq.
20]{roxburgh_asteroseismic_2016}: \begin{equation}
    r_{02}(\nu_{n,0}) = {\nu_{n,0} - \nu_{n-1,2} \over \nu_{n,2} - \nu_{n - 1,2}} \sim {1\over\pi}\left[\delta_{2, \pi}(\nu_{n,0}) - \delta_{0, p}(\nu_{n,0})\right].
\end{equation} As before, these are treated as functions of frequency,
with model quantities interpolated to the observed radial mode
frequencies for the purposes of computing cost functions. While
isochrones on the \(r_{02}\)-\Dnu~diagram (with this modified
definition) have been shown to yield good constraints on main-sequence
stellar ages in numerical studies \citep{white_calculating_2011}, the
behaviour of these ratios for evolved stars is largely unexplored. We
include it for completeness in comparison with our main-sequence sample.

This issue of mode mixing also afflicts the phase-matching algorithm,
where the constructed phase differences collapse to a single function
only when comparing two sets of \(\pi\)-modes (or uncoupled p-modes).
Again, we exclude dipole modes from the phase-matching computation for
red giants.

Finally, while mixed-mode resonances resulting in avoided crossings
should in principle be determined strictly after the application of the
surface correction to the uncoupled \(\pi\)-mode frequencies, the
\bg~correction is typically applied directly to the model mixed modes
instead (i.e.~after, and not before, accounting for coupling to
g-modes). \citet{ball_surface_2018} demonstrate that these two
operations do not commute. Nonetheless, they also show that the
parameterisation of \bg~remains sufficiently applicable to the uncoupled
\(\pi\)-mode frequencies to permit their use in stellar modelling as
well, assuming again that these \(\pi\)-modes can be accurately
estimated. For this reason, and for consistency with both of our
nonparametric methods, we again only use quadrupole modes for analysing
these red giants.

\hypertarget{results}{%
\subsection{Results}\label{results}}

\begin{figure}[htbp]
    \centering
    \includegraphics[width=.46\textwidth, trim=.15cm .75cm .25cm .15cm,clip]{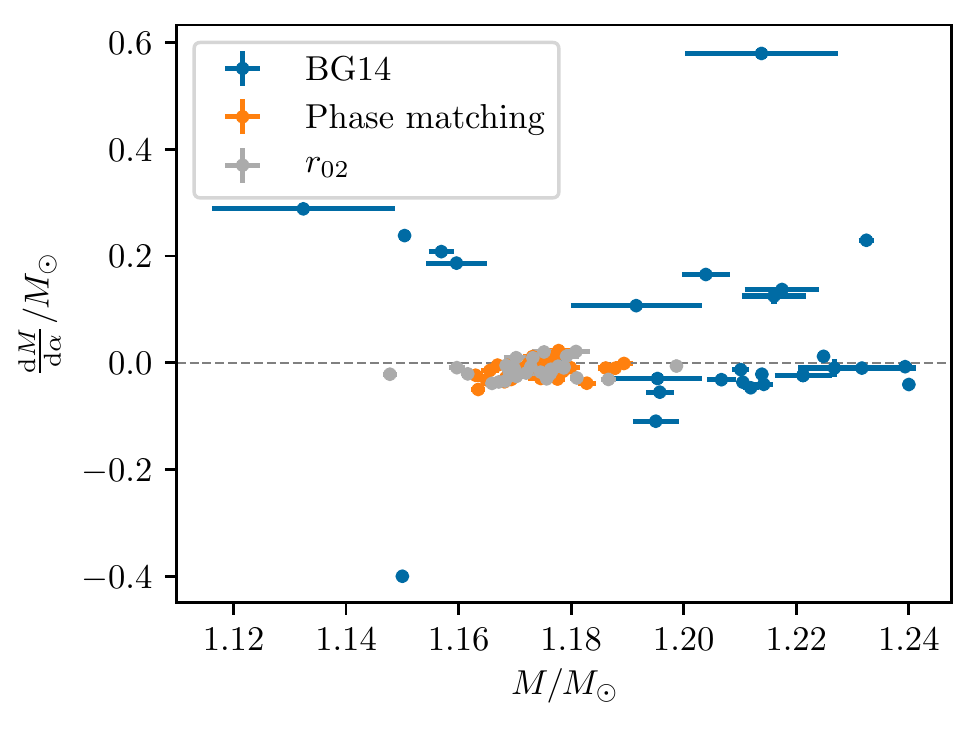}
    \includegraphics[width=.45\textwidth, trim=.15cm .25cm .25cm .15cm,clip]{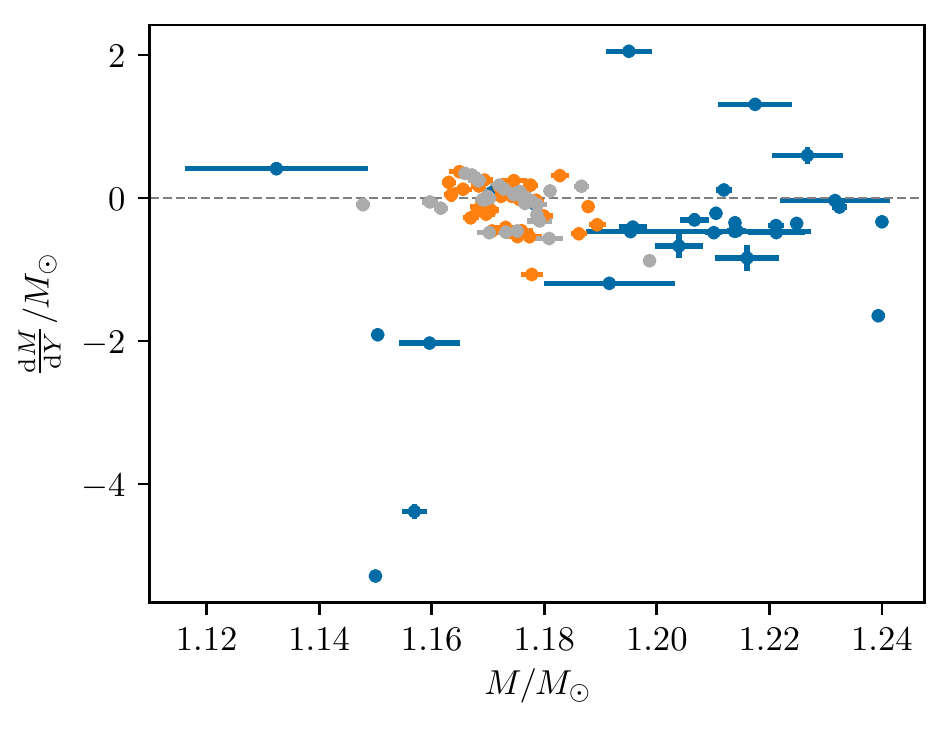}
    \caption{Correlation parameters, as in \cref{fig:ms}, for the posterior mass vs. \amlt\ (top) and $Y_0$ (bottom) for red giants in NGC\ 6791.}
    \label{fig:rg}
\end{figure}

Unlike the case of main-sequence stars, we find that different choices
of surface correction yield qualitatively different parameter
inferences, even to leading order. We show in \cref{fig:rg} the
distribution of inferred masses, and corresponding correlation
parameters, as done in \cref{fig:ms}. It is clear that both of the
nonparametric algorithms return a considerably smaller spread in the
inferred stellar masses than does the \bg~correction, although on
average the pairwise normalised differences do not differ very
significantly from zero (as listed in \cref{tab:rg}). Moreover, the
results from both nonparametric surface treatments are highly consistent
with each other. This is also the case for other inferred parameters,
such as the initial helium abundances and stellar ages.

\begin{longtable}[]{@{}lrccc@{}}
\caption{\(t\)-test \(p\)-values for normalised differences between
parameters inferred under different surface corrections for
NGC~6791.\label{tab:rg}}\tabularnewline
\toprule
Quantity & \bg~vs.~phase & \bg~vs.~\(r_{02}\) & phase vs.~\(r_{02}\)
&\tabularnewline
\midrule
\endfirsthead
\toprule
Quantity & \bg~vs.~phase & \bg~vs.~\(r_{02}\) & phase vs.~\(r_{02}\)
&\tabularnewline
\midrule
\endhead
Mass & \(0.0171\) & \(0.0025\) & \(0.8215\) &\tabularnewline
Radius & \(0.1005\) & \(0.0020\) & \(0.0264\) &\tabularnewline
Age & \(0.0016\) & \(0.0010\) & \(0.9770\) &\tabularnewline
\(Y_0\) & \(0.0345\) & \(0.0275\) & \(0.3983\) &\tabularnewline
\bottomrule
\end{longtable}

In all of these cases, we have modelled each star separately, and so
have not constrained the masses to be similar to each other.
Nonetheless, seeing as all of these stars are from the same open
cluster, all first-ascent red giants should be of essentially the same
age and composition, and therefore mass. Consequently, we should prefer
a smaller spread in the stellar masses a priori. It is therefore quite
astrophysically significant that nonparametric treatments of the surface
term should return so much less scatter than the \bg~correction. The
most likely reason for this is that the \bg~correction permits far more
flexibility than strictly required to describe the surface term in such
evolved stars: while the surface properties of main-sequence stars may
exhibit significant differences as a matter of mass, age, and
composition, all red giants are essentially similar (being almost fully
convective).

This result is qualitatively different from that of
\citet{jorgensen_investigating_2020}, who studied different parametric
surface corrections for the open clusters NGC~6791 and NGC~6819, where
it was found that various parametric surface corrections all yielded
comparable scatter in the inferred masses for both clusters. We note
that our observed scatter in the inferred masses from the \bg~correction
is consistent with previous studies of NGC~6791
\citep{mckeever_helium_2019, jorgensen_investigating_2020} using the
same \bg~surface correction. In particular, it is consistent with with
results from the grid used in \citet{jorgensen_investigating_2020},
which was computed over a considerably larger range of stellar masses
(0.7 to 2.5 \(M_\odot\)) and used a different set of physical
prescriptions. In our analysis of nonparametric treatments of the
surface term, we have taken care to hold constant all other properties
of both the grid and the inference methodology. Therefore, we can
attribute the markedly reduced scatter in the inferred masses in the
other cases solely to the choices of (nonparametric) surface correction.
Corollarily, this may suggest that parametric treatments of the surface
term, in general, may potentially overfit frequency differences in
evolved stars.

Finally, \cref{fig:rg} also shows that the dependence of the inferred
masses on the model parameters \amlt~and \(Y_0\) is also differently
distributed (on the vertical axes) between the three surface corrections
we have considered. Unfortunately, as we have noted earlier, the
discretised nature of the model grid of \citet{mckeever_helium_2019}
does not permit us to examine the effects of more sophisticated
constraints on model selection, as we did with the LEGACY sample.

\hypertarget{discussion-and-conclusion}{%
\section{Discussion and Conclusion}\label{discussion-and-conclusion}}

Asteroseismic analysis requires that the surface term be accounted for
when using detailed mode frequencies as a modelling constraint; there
are many competing proposals for how this is to be done. Previous works
have established that the inferred values of many stellar parameters
using asteroseismic inputs are robust to different choices of parametric
seismic surface corrections, both on the main sequence
\citep{compton_surface_2018, nsamba_asteroseismic_2018} and for evolved
stars
\citep{ball_surface_2017, ball_surface_2018, jorgensen_investigating_2020}.
On the other hand, there exists a different class of surface
corrections, which are nonparametric and motivated from theoretical
rather than phenomenological considerations.
\citet{nsamba_asteroseismic_2018} shows the method of separation ratios,
in particular, to be in agreement with other parametric corrections for
some Kepler main-sequence stars, while \citet{basu_robustness_2018}
demonstrated this to also be the case for the phase-matching method as
well, albeit on a small sample. We show that these results continue to
hold, in a statistical sense, on the full Kepler LEGACY sample. The
exception to this is the initial helium abundance, where one of the
surface corrections we have examined, the method of phase matching,
yields statistically significant differences from the more commonly
employed parametric corrections. This suggests that presently ongoing
efforts to constrain helium abundances from TESS asteroseismology will
be sensitive to the choice of surface correction, which should now be a
new methodological consideration for these and future studies (e.g.~for
the PLATO mission). On the other hand, the phase-matching method in
particular is not commonly used for stellar modelling, precisely because
of its slightly more computationally expensive nonparametric structure.
Further investigation will be necessary to determine if this additional
investment will be worth the expense.

Other mixing processes, such as core over/undershoot, are often also
invoked as additional free parameters in detailed seismic modelling in
order to yield better structural matches, particularly with stars
exhibiting convective cores. Although we do not examine the effects of
varying these in this work, we are able to demonstrate, in a restricted
sense, that their interaction with the surface term is minor, and that
the different surface-term prescriptions form behave similarly (see
\autoref{sec:overshoot}).

We have also shown that the situation is drastically different for red
giant stars in our NGC~6791 sample, where nonparametric treatments of
the surface term yield qualitatively different mass estimates compared
to the parametric \bg~correction: inferences for individual stars appear
to be more precise, and we also obtain far less statistical scatter when
considering the sample of parameters returned from modelling each star
in a open cluster separately. To our knowledge, this is the first
comparison between parametric and nonparametric treatments of the
surface term for evolved stars. It is tempting to conclude from this
that such methods should be preferred when modelling these evolved
stars. However, this would be a hasty generalisation, given that we have
only studied a single open cluster. Further investigation of this
nature, repeating this analysis for red giants in other open clusters
(e.g.~NGC~6819) for which detailed asteroseismology is also available,
must be done before any general conclusions about red giants may be
drawn.

The phase-matching and separation-ratio methods are not commonly applied
to red giants. While we have demonstrated their viability as applied to
only quadrupole mixed modes (assumed to be weakly coupled to the inner
g-mode cavity), more theoretical work may be required to also include
dipole modes in the constraints, where the mixing with the g-mode cavity
is known to have a more pronounced effect (see \autoref{ap:coupling}).

\acknowledgements

We thank B. Mosser and B. Nsamba for productive discussions, and we
thank the anonymous referee for helpful comments and suggestions. JO and
JMM thank the Yale Center for Research Computing for guidance and use of
the research computing infrastructure.

\software{NumPy \citep{numpy}, SciPy stack \citep{scipy}, AstroPy \citep{astropy:2013,astropy:2018}, Pandas \citep{mckinney-proc-scipy-2010}, \mesa\ \citep{mesa_paper_1,mesa_paper_2,mesa_paper_4}, \gyre\ \citep{townsend_gyre_2013}, \texttt{corner} \citep{dfm_corner_2016}}

\facilities{Kepler}

\appendix

\hypertarget{relative-coupling-strengths-of-dipole-vs.-quadrupole-mixed-modes}{%
\section{Relative coupling strengths of dipole vs.~quadrupole mixed
modes}\label{relative-coupling-strengths-of-dipole-vs.-quadrupole-mixed-modes}}

\label{ap:coupling}

We wish to demonstrate that, at least for our red giant sample, the
\(\pi\)-\(\gamma\) coupling for quadrupole modes is sufficiently weak
that \(r_{02}\) and \(\delta_{\pi, 2}(\nu)\) may be meaningfully
estimated from the observed p-dominated mixed modes. To do this, we will
need the following facts:

\begin{enumerate}
\def\labelenumi{\arabic{enumi}.}
\tightlist
\item
  The red giants in our NGC~6791 sample are all close to the RGB bump.
\item
  Mode coupling between the \(\pi\) and \(\gamma\) cavities admits a
  functional description with coupling and norm matrices that yield a
  Hermitian eigenvalue problem, which we write in block matrix form
  as\begin{equation}
   \mathbf{L} = \begin{bmatrix}
   \mathbf{L}_\pi & \mathbf{R} \\ \mathbf{R} & \mathbf{L}_\gamma
   \end{bmatrix} = -\omega^2 \mathbf{D},
  \end{equation} with explicit expressions for these block matrix
  elements given in \citet{ong_semianalytic_2020} as integrals against
  the isolated eigenfunctions. In particular, let us consider the
  coupling between a single \(\pi\) mode and a single \(\gamma\) mode,
  so that this eigenvalue problem takes the form \begin{equation}
   \begin{bmatrix}
   -\omega_p^2 & \alpha - D \omega_p^2 \\ \alpha - D \omega_p^2 & -\omega_g^2
   \end{bmatrix} = -\omega^2 \begin{bmatrix}
   1 & D \\ D & 1
   \end{bmatrix},
  \end{equation} with eigenvalues \(\omega^2_+, \omega^2_-\), where
  \(\alpha\) is the \(\pi\)-\(\gamma\) coupling strength and \(D\) is
  the volume overlap integral of their eigenfunctions. Without loss of
  generality, we choose a normalisation of the eigenfunctions so that
  both are positive. It can be shown that the perturbation to the
  eigenvalues with respect to the uncoupled values
  (e.g.~\(\delta \omega^2_{\pi\gamma} = \omega^2_+ - \max(\omega^2_p, \omega^2_g)\))
  is largest at resonance, where \(\omega_p = \omega_g = \omega\),
  yielding \(\delta \omega^2_\text{upper} = {\alpha \over 1 - D}\). This
  is a strict upper bound, so that in general
  \(|\delta\omega^2_{\pi\gamma}| < \delta\omega^2_\text{upper}\). In
  turn, this yields an upper bound on the mode bumping as
  \(|\delta\nu_{\pi\gamma}| \lesssim {\nu \over 2}\left(\delta \omega^2_\text{upper} \over \omega^2\right) \equiv \delta\nu_\text{upper}\).
\item
  In practice, the true mode bumping for any given mode will on average
  also be smaller than this upper bound, since the \(\pi\) and
  \(\gamma\) modes evolve monotonically in different directions along
  the RGB, and so do not stay in resonance for long. Only modes whose
  eigenvalues are sufficiently close, as defined by a characteristic
  pairwise resonance width
  (i.e.~\(|\omega_p^2 - \omega_g^2| < \Gamma^2\)), will have their
  frequencies significantly perturbed through an avoided crossing. This
  resonance width \(\Gamma\) also scales with
  \(\delta\omega_\text{upper}\).
\end{enumerate}

\begin{figure}
\centering
\includegraphics{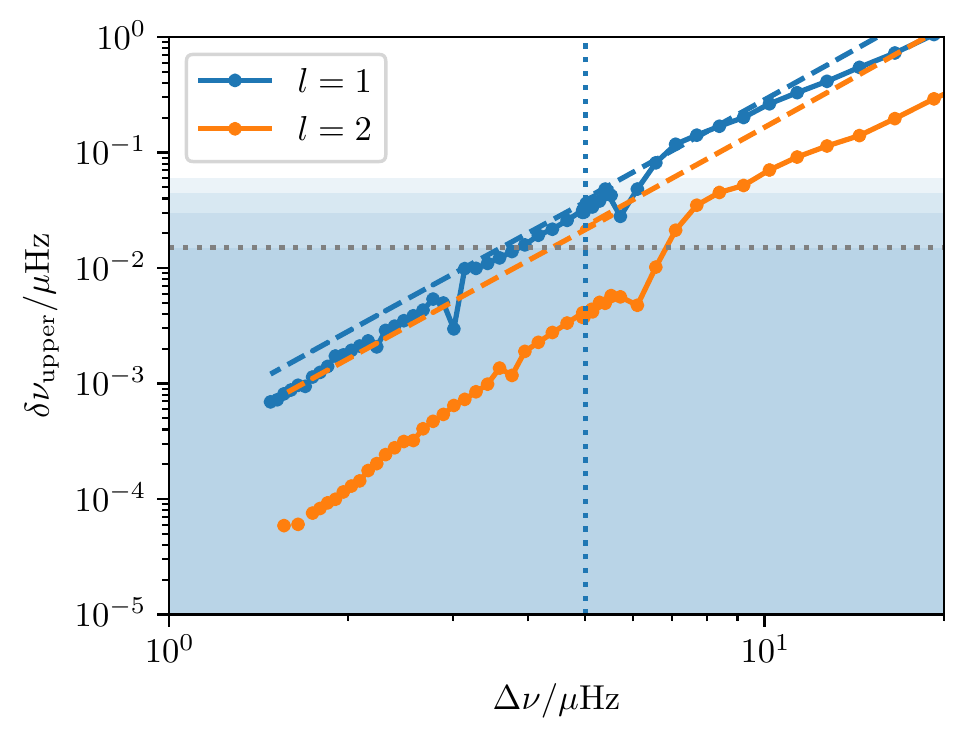}
\caption{Evolution of the upper bound \(\delta\nu_\text{upper}\) for the
frequency perturbation resulting from mode bumping. See main text for a
full description. \label{fig:bump}}
\end{figure}

We plot this quantity in \cref{fig:bump} as a function of \(\Delta\nu\)
for the first RGB ascent of a 1.18 \(M_\odot\) MESA evolutionary track
below, for the dipole and quadrupole modes closest to
\(\nu_\text{max}\). We mark the location of the RGB bump with a vertical
dotted line, and we mark out a representative value of the mode
frequency \(1\sigma\) measurement error with a horizontal dotted line,
with shaded regions for integer multiples thereof to guide the eye.
Finally, we mark out the local \(\gamma\)-mode frequency spacing
\(\nu^2 \Delta\Pi_l\) evaluated at \(\nu_\text{max}\), with dashed
lines.

With respect to this diagram, we make the following observations:

\begin{itemize}
\tightlist
\item
  The coupling strength, and so the maximum possible frequency
  perturbation from mode bumping via the avoided crossing phenomenon, is
  an order of magnitude weaker for quadrupole modes than dipole modes.
\item
  The resonance widths for dipole modes are comparable to the
  \(\gamma\)-mode spacing, but for quadrupole modes they are much
  smaller than the \(\gamma\)-mode spacing. Consequently, whereas dipole
  \(\pi\) modes near \(\nu_\text{max}\) will be close to resonance with
  at least one \(\gamma\) mode at any given time, a quadrupole \(\pi\)
  mode must be fortuitiously in resonance with the underlying
  \(\gamma\)-modes, in order for the most p-dominated mixed mode
  corresponding to it to have its frequency significantly differ from
  the uncoupled value.
\item
  Even were this to be the case, the magnitude of this frequency
  perturbation is bounded from above. For the stars in our RGB sample
  (\(\Delta\nu \lesssim 8\ \mu\)Hz), the quadrupole upper bound is
  smaller than, or at most comparable to, the frequency measurement
  error. That is to say, for the majority of these stars, the
  observational data cannot distinguish between the uncoupled quadrupole
  \(\pi\) modes and the worst-case bumped frequencies of the most
  p-dominated quadrupole mixed modes.
\end{itemize}

\hypertarget{sensitivity-to-overshoot}{%
\section{Sensitivity to overshoot}\label{sensitivity-to-overshoot}}

\label{sec:overshoot}

In this work, we have restricted the choice of physics used in the
stellar evolution, since our interest does not lie in accurately
estimating stellar properties per se, but rather in isolating the
effects of changing the input physics from those of using a different
surface correction prescription. For this purpose we have adopted a
uniform set of model physics, so that any differences in the inferred
properties arise solely from differences in the surface corrections.

\begin{figure}
\centering
\includegraphics{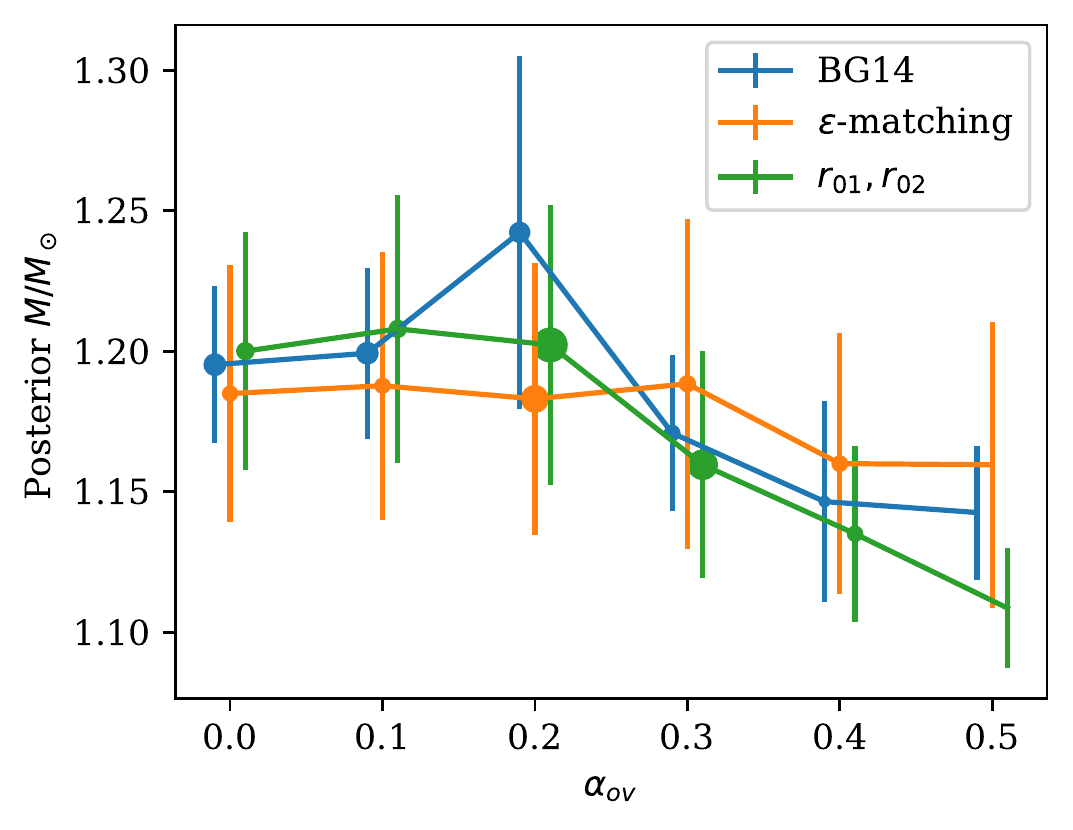}
\caption{Inferred conditional posterior masses of KIC 12258514 as a
function of \(\alpha_\text{ov}\). \label{fig:overshoot}}
\end{figure}

Nonetheless, we wish to demonstrate that surface-term-driven systematic
effects are largely separable from those arising from different
treatments of core overshoot in particular. For this demonstration, we
will consider the single star KIC 12258514, which is a LEGACY star that
is know to have a convective core. We use the model grid of
\citet{viani_overshoot_2020}, which was constructed with YREC
\citep{demarque_yrec_2008}. The overshoot parameter \(\alpha_\text{ov}\)
was allowed to vary freely in this grid, and was discretely sampled. We
apply the modelling procedure of that work --- i.e.~likelihood-weighted
averaging --- appropriately modified to include the regularisation term
and different surface corrections used in this work, to each of these
different subgrids, corresponding to different values of
\(\alpha_\text{ov}\). We show the conditional posterior masses resulting
from this procedure in \autoref{fig:overshoot}. The data points are
offset horizontally by surface correction to permit easy visual
comparison, and are sized by the maximum of the likelihood function for
each subgrid (larger points indicate lower minimum \(\chi^2\), which are
more likely).

With respect to this figure, we observe the following:

\begin{itemize}
\tightlist
\item
  For each choice of surface correction, the overall variation in the
  conditional posterior masses (as the overshoot parameter
  \(\alpha_\text{ov}\) is varied) is in any case comparable to the
  inferred posterior mass error.
\item
  For each adopted value of the overshoot parameter, the variation in
  posterior mass between surface corrections is also comparable to the
  inferred posterior mass error.
\end{itemize}

\citet{viani_overshoot_2020} created independent model grids for each
star, making it, unfortunately, difficult to do a population-wide
comparison of the kind that we have sought to do in this work. While
this result is not necessarily representative of all stars in the
sample, it serves to demonstrate that any interaction between surface
corrections and overshoot is likely to be small for stars considered
individually \citep[as
in][]{compton_surface_2018, basu_robustness_2018}, and hence that these
parameter inferences are likely robust to the surface term even when
overshoot is varied.

  \bibliography{biblio.bib}

\end{document}